\documentclass[12pt]{article}
\usepackage[dvips]{graphicx}
\usepackage{epsfig}
\usepackage{amssymb} %\graphicspath{{Figures_L/}}

\def\mev{\,{\rm MeV}}
\def\gev{\,{\rm GeV}}

\newcommand{\nn}{\nonumber}
\newcommand{\be}{\begin{equation}}
\newcommand{\ee}{\end{equation}}
\newcommand{\ba}{\begin{eqnarray}}
\newcommand{\ea}{\end{eqnarray}}
\newcommand{\lsim}{\raisebox{-4pt}{$\,\stackrel{\textstyle
                                        <}{\sim}\,$}}
\newcommand{\tw}{\textwidth}

\def\={\,=\,}
\newcommand{\gsim}{\raisebox{-4pt}{$\,\stackrel{\textstyle
                                            >}{\sim}\,$}}
\newcommand{\req}[1]{(\ref{#1})}

\newcommand{\tr}[1]{ {\bf #1}_\perp}
\newcommand{\ov}[1]{\overline#1}

\def\cbc{{c}\bar{c}}
\def\ubu{{u}\bar{u}}

\def\als{\alpha_s}

\def\vd{{\bf \Delta}_\perp}
\def\vdd{ \Delta_\perp^2}
\newcommand{\vk}{{\bf k}_{\perp}}
\def\vkk{k^2_\perp}

\def\wf{wave function}
\def\ci{\cite}
\def\sh{\hat{s}}
\def\uh{\hat{u}}
\def\th{\hat{t}}
\def\xb{\bar{x}}

\begin{document}
\thispagestyle{empty}
\begin{flushright}
WU B 09-04 \\
hep-ph/0905.2561\\
May 2009\\[5em]
\end{flushright}

\begin{center}
{\Large\bf Proton-Antiproton Annihilation into
a $\Lambda_c^+ \bar{\Lambda}{}_c^-$ Pair} \\
\vskip 3\baselineskip

A.T.\ Goritschnig$^{a}$\,\footnote{Email:
alexander.goritschnig@uni-graz.at} P.\
Kroll$^{b,c}$\,\footnote{Email:
kroll@theorie.physik.uni-wuppertal.de}
W.\ Schweiger$^{a}$\,\footnote{Email: wolfgang.schweiger@uni-graz.at}\\[0.5em]
a) {\small {\it Institut f\"ur Physik, Universit\"at Graz,
    8010 Graz, Austria}}\\
b) {\small {\it Fachbereich Physik, Universit\"at Wuppertal,
 42097
Wuppertal, Germany}}\\
c) {\small {\it Institut f\"ur Theoretische Physik, Universit\"at
    Regensburg,   93040 Regensburg, Germany}}\\
\end{center}
\vskip \baselineskip
\begin{center}
\begin{abstract}
The process $p\bar{p}\to\Lambda_c^+ \bar{\Lambda}{}_c^-$ is
investigated within the handbag approach. It is shown that the
dominant dynamical mechanism, characterized by the partonic
subprocess $u\bar{u}\to c\bar{c}$, factorizes in the sense that
only the subprocess contains highly virtual partons, a gluon to
lowest order of perturbative QCD, while the hadronic matrix
elements embody only soft scales and can be parameterized in terms
of helicity flip and non-flip generalized parton distributions.
Modelling these parton distributions by overlaps of light-cone \wf
s for the involved baryons we are able to predict cross sections
and spin correlation parameters for the process of interest.
\end{abstract}
\end{center}

%%%%%%%%%%%%%%%%%%%%%%%%%%%%%%%%%%%%%%%%%%%%%%%%%%%%%%%%%%%%%%%%%%%%%%
\section{Introduction}
%%%%%%%%%%%%%%%%%%%%%%%%%%%%%%%%%%%%%%%%%%%%%%%%%%%%%%%%%%%%%%%%%%%%%%
The FAIR project at GSI with the HESR antiproton program will
offer ideal possibilities to study exclusive channels in
$p\bar{p}$ annihilation. The energy of the antiproton beam
suffices to produce pairs of heavy baryons, as for instance
$\Lambda_c^+\bar{\Lambda}{}_c^-$, in proton-antiproton collisions.
Prerequisite for the measurement of such processes is, however,
that their cross sections are sufficiently large. The purpose of
this work is to estimate the cross section for the reaction
$p\bar{p}\to\Lambda_c^+\bar{\Lambda}{}_c^-$. This estimate is
based on a generalization of the, so-called, \lq\lq handbag
approach\rq\rq,  which has been developed for deeply virtual
electroproduction of photons and mesons \ci{mueller94}. It has
been firmly shown that the amplitudes for these processes are
described by convolutions of hard subprocess kernels and,
so-called, \lq\lq generalized parton distributions\rq\rq\ (GPDs),
which encode the soft, non-perturbative physics. A second class of
hard exclusive reactions to which the handbag approach can be
applied to, is formed by wide-angle reactions. The hard scale for
these reactions, necessary to justify factorization  into hard and
soft physics, is provided by the momentum transfer from the
incoming to the outgoing hadron or, in other words, by large
Mandelstam variables $-t$ and $-u$, instead of the photon
virtuality. Although, in contrast to deeply virtual processes,
rigorous proofs do not exist, arguments for the dominance of the
handbag contribution have been given in some cases
\ci{DFJK1,rad,huang02}.

Along the lines of argumentation for wide-angle Compton scattering
\ci{DFJK1} we demonstrate that, under the assumption of restricted
parton virtualities and transverse momenta, the
$p\bar{p}\to\Lambda_c^+\bar{\Lambda}{}_c^-$  amplitudes factorize
into hard subprocesses and moments of the $u\!\to\!c$ transition
GPDs. Owing to the relatively heavy charm quarks (with  mass
$m_c$), the process possesses the large intrinsic scale
$4m_c^2\simeq 6.3\,\gev^2$ to which the momentum transfer adds.
The large intrinsic scale leads to the highly welcome consequence
that large $-t$ and $-u$ are not required as, say, for wide-angle
Compton scattering. Therefore the arguments for factorization of
the $p\bar{p}\to\Lambda_c^+\bar{\Lambda}{}_c^-$  amplitude already
hold for small scattering angles.

It is to be stressed that the process of interest has already been investigated
long time ago \ci{quadder} in an approach that bears resemblance to the
handbag one. As compared to \ci{quadder} the progress achieved here manifests
itself in the more solid theoretical basis. The normalization of the cross
section is now determined for given GPDs and not fixed by symmetry arguments as
in \ci{quadder}. Moreover, the complicated structure of the proton which is
described in terms of eight GPDs, can be easily taken into account in the
handbag approach and therefore the spin dependence of the process
$p\bar{p}\to\Lambda_c^+\bar{\Lambda}{}_c^-$ can readily be treated.

The plan of the paper is the following: The arguments for factorization and
the principle structure of the handbag contribution are presented in
Sect.\ \ref{sec:factorization}. In Sect.\ \ref{sec:amplitudes} the amplitudes
of the process $p\bar{p}\to\Lambda_c^+\bar{\Lambda}{}_c^-$ are discussed while the
subprocess amplitudes are given in Sect.\ \ref{sec:subprocess}. Various models
for the GPDs are presented in Sect.\ \ref{sec:GPD} and predictions for
observables are discussed in Sect.\ \ref{sec:observables}. The summary is
given in Sect.\ \ref{sec:summary}. In two appendices various kinematical
formulas and the properties of the transition GPDs are compiled.

%%%%%%%%%%%%%%%%%%%%%%%%%%%%%%%%%%%%%%%%%%%%%%%%%%%%%%%%%%%%%%%%%%%%%%
\section{Factorization} \label{sec:factorization}
%%%%%%%%%%%%%%%%%%%%%%%%%%%%%%%%%%%%%%%%%%%%%%%%%%%%%%%%%%%%%%%%%%%%%

%%%%%%%%%%%%%%%%%%%%%%%%%%%%%%%%%%%%%%%%%%%%%%%%%%%%%%%%%%%%%%%%%%%
\subsection{The basic idea}
\label{sec:basic-idea}
%%%%%%%%%%%%%%%%%%%%%%%%%%%%%%%%%%%%%%%%%%%%%%%%%%%%%%%%%%%%%%%%%%%
\begin{figure}[t]
\begin{center}
\includegraphics[width=.47\textwidth,bb=115 455 408 643,%
clip=true]{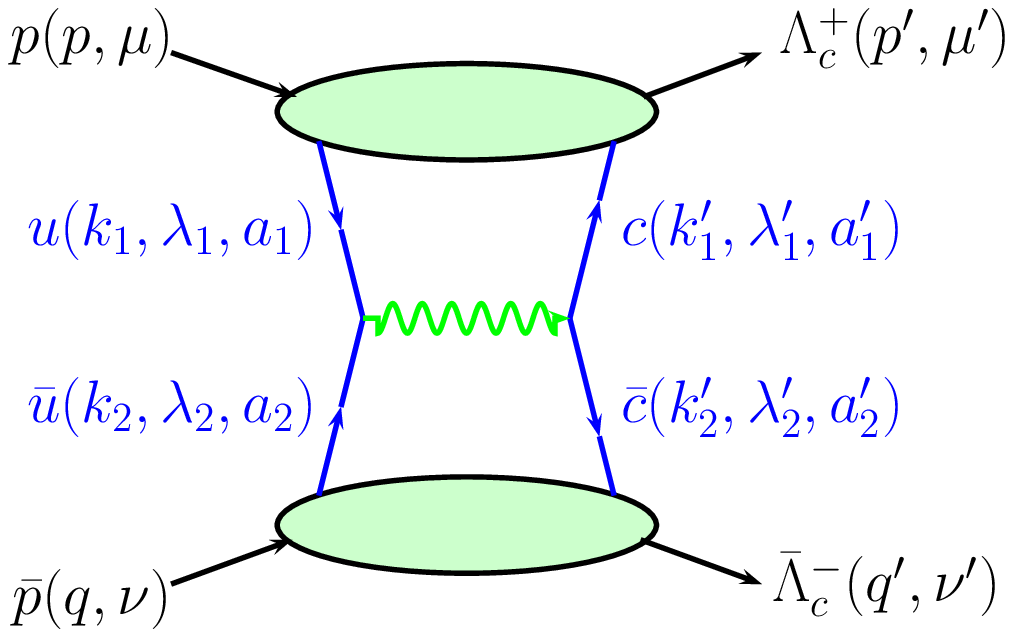}
\end{center}
\caption{The handbag contribution to the process $p\bar{p}\to
  \Lambda_c^+ \bar{\Lambda}_c^-$. The momenta and helicities of the
  baryons and quarks as well as the colors of the latter, are also specified.}
\label{fig:handbag}
\end{figure}
We assume that the graph shown in Fig.\ \ref{fig:handbag}
dominates the process of interest for Mandelstam $s$ well above
the kinematical threshold and in the forward scattering
hemisphere. We are going to show in this section that the  baryons
emit and re-absorb soft partons, i.e.\ quarks that are almost
on-shell and move almost collinear with their parent baryon. In
other words, the contribution from the graph shown in Fig.\
\ref{fig:handbag} factorizes into a hard partonic subprocess and
soft hadronic matrix elements. The arguments for factorization go
along the same lines as for wide-angle Compton scattering
\cite{DFJK1}, except that here we have to consider the double
handbag and we have to take care of the unequal mass
kinematics~\footnote{
 With the exception of the latter issue this is similar to proton-proton
elastic scattering which has been briefly discussed in
\cite{DFJK2}.}. For the $p \to \Lambda_c^+$ transition a $u$-quark
is to be emitted and a $c$-quark reabsorbed; the valence quark
content is to be changed. The emission of a light quark other than
the $u$ one from the proton does not lead to a $\Lambda_c$ baryon
in the final state. Intrinsic charm would allow for subprocesses
like
\be \bar{c}\, \bar{u}\,  \to \,\bar{u} \,\bar{c}\,.
\ee
This is a higher Fock-state contribution which is expected to be
suppressed by general arguments. Moreover, according to present
PDF analyses, e.g.\ \ci{CTEQ6,MRST04}, the charm content of the
proton is tiny. Even at a scale of, say, $25\,\gev^2$ it only
amounts to about $10^{-3}$ of the $u$-quark content at $x\simeq
0.6$. Thus, we merely have to take into account the subprocess
\be u \bar{u}  \to c \bar{c}\,. \label{subprocess} \ee
This subprocess goes along with $p\to \Lambda_c^+$ transition
matrix elements which, as we are going to show below, can be
parameterized by $u\to c$ GPDs. One may also think of time-like
matrix elements or GPDs representing $p\bar{p}\to \ubu$ and
$\cbc\to\Lambda_c^+ \bar{\Lambda}{}_c^-$ transitions. Such
time-like GPDs have been employed in the analysis of wide-angle
time-like processes as, for instance, two-photon annihilation into
a pair of hadrons \ci{DKV2,DKV3,weiss}. To lowest order of QCD the
subprocess is mediated by one-gluon exchange which evidently
forbids the time-like reaction mechanism by color conservation to
this order of accuracy. It is, however, allowed to higher orders
of QCD although, beyond representing an $\als$ correction,
expected to be smaller than the mechanism depicted in Fig.\
\ref{fig:handbag} since the $p\to \Lambda_c^+$ transitions seem to
favored over the time-like ones. We therefore do not consider the
time-like mechanism.

A $u\to c$ transition GPD is a function of three variables, a
momentum fraction $\xb$ defined by the ratio of light-cone plus
components of the average parton and baryon momenta (see Fig.\
\ref{fig:handbag})
\be \xb_1 \= \frac{k_1^+ + k_1^{\prime +}}{p^+ + p^{\prime +}}\,,
\ee the skewness parameter \be \xi \= \frac{p^+ - p^{\prime
+}}{p^+ + p^{\prime +}}\,, \ee
and Mandelstam $t$. Analogous definitions hold for the antibaryon
vertex. The $u\to c$ transition GPDs are expected to exhibit a
maximum which becomes more pronounced as the mass of the heavy
quark increases. The position of the peak is approximately at
\be x_0\=m_c/M \= 1-\epsilon/M\,, \label{peak-position} \ee
for not too large $-t$. The width of the peak shrinks with
increasing mass. For standard values of the charm quark mass
\ci{PDG}
\be m_c\=(1.27 \textstyle{+ 0.07 \atop -0.11})\,\gev\,,
\label{c-error} \ee
$x_0$ is about 0.5 - 0.6. The parameter $\epsilon$ is the
difference between the hadron and the heavy quark mass and has a
value of about $0.6 - 1.0\,\gev$. In the formal limit of $m_c\to
\infty$ $x_0$ tends to 1 according to the heavy quark effective
theory (HQET)~\ci{HQET}. This expected property of the transition
GPDs parallels the theoretically expected and experimentally
confirmed behavior of heavy quark fragmentation functions, in
particular $c\to \Lambda_c^+$ (see \cite{kniehl05,kniehl06}) and
is also analogous to the behavior of light-cone \wf s (LCWFs) and
distribution amplitudes for heavy baryons
\cite{koerner,ball-braun}). The $x$ dependence of the transition
GPD is to be contrasted with wide-angle Compton scattering
\cite{DFJK1} and the zero-skewness GPD analysis \cite{DFJK4} where
the GPDs exhibit a behavior similar to the  one expected for the
$u\to c$ transition only for $-t\gg 1\gev^2$.

Some kinematical details of our process of interest are compiled
in App.\ A. As for Compton scattering we work in a symmetric frame
with the baryon momenta as specified in \req{def-momenta-ji} and
assume that the virtualities of all the partons the baryons are
made of are smaller then $\Lambda^2$ where $\Lambda$ is a typical
hadronic scale of order $1\,\gev$. Of course, the virtualities of
the charm and anticharm quark respect  the relation $\mid
k^{\prime 2}-m_c^2\mid \leq \Lambda^2$. We further assume that all
parton transverse momenta satisfy the restrictions:
\be \tilde\vk^2{}_i/\tilde{x}_i\lsim \Lambda^2\,, \qquad
\hat\vk^{\prime\, 2}{}_i/\hat{x}^\prime_i \lsim \Lambda^2\,,
\label{assumption-1} \ee
where, for the ease of legibility, we
have not distinguished between the parton momenta from the upper
and lower vertex. The bounds for the transverse momenta are
assumed to hold in hadron frames in which the respective hadron
moves along the 3-direction (the tilde characterizes momenta in
the hadron-in frame, the hat those in the hadron-out frame). These
hadron frames are reached by transverse boosts, cf.\ e.g.\
\cite{brodsky89}. This is a transformation that leaves the plus
component of any momentum vector $a$ unchanged. It involves a
parameter $b^+$ and a transverse vector $\tr{b}$ and is defined by
\be
 [a^+,a^-,\tr{a}]  \longrightarrow
   \left[a^+,a^- - \frac{\tr{a}\cdot\tr{b}}{b^+}+ \frac{a^+ \tr{b}^2}{2(b^+)^2},
                                   \tr{a}-\frac{a^+}{b^+}\tr{b}\right]\,.
\label{trans-boost} \ee
The corresponding transformation for antibaryons requires
transverse boosts that leave the minus components of the momenta
unchanged. In the following we discuss the baryon vertex in
detail. All the arguments and results given below do hold for the
antibaryon vertex analogously.

The active parton momenta are parameterized as
\be k_1 \= [x_1{p}^+,k_1^-,{\bf k}_{1\perp}]\,, \qquad
k_1^\prime\=[x_1^\prime {p}^{\prime +},k_1^{\prime -}, {\bf
k}_{1\perp}^\prime]\,. \label{parton-momenta} \ee
The individual momentum fractions $x_1$ and $x_1^\prime$ are
related to $\xb_1$ and the skewness by
\be x_1\= \frac{\xb_1+\xi}{1+\xi}\,, \qquad x_1^\prime\=
\frac{\xb_1-\xi}{1-\xi}\,. \label{individual-fractions} \ee
Transforming the parton momenta \req{parton-momenta} to the
respective hadron-in or hadron-out frame by the transverse boost
\req{trans-boost}, we find
\be \tilde{k}_1\= [x_1{p}^+,\ldots,{\bf k}_{1\perp}+x_1\vd/2]\,,
\qquad \hat{k}_1^\prime\= [x_1^\prime{p}^{\prime +},\ldots, {\bf
k}_{1\perp}^\prime-x_1^\prime\vd/2]\,. \ee
For convenience let us assume that there is only one
spectator~\footnote{ This may be viewed as a quark-diquark
configuration of the baryons.}. The generalization to an arbitrary
number of spectators is straightforward but tedious. As is
characteristic of any parton approach the momenta of the active
parton ($k_1$ and $k_1^\prime$) and the spectator ($k_s$ and
$k_s^\prime$) sum up to the parent baryon's momentum. Hence,
\be \vk{}_s\=-\vd/2-{\bf k}_{1\perp}\,, \qquad x_s\=1-x_1\,, \ee
and analogously for $k^\prime_s$. The spectator condition
$k^{\phantom{\prime}}_s=k^\prime_s$ leads to
$x_1^\prime(1-\xi)=x_1(1+\xi) -2\xi$ and ${\bf k}_{1\perp}^\prime
=\vd+{\bf k}_{1\perp}$. Transforming also the spectator momenta to
the respective hadron frames and using the supposition
\req{assumption-1}, we can write
\be 2\Lambda^2 \geq \tilde{k}^2_{\perp s}/\tilde{x}_s +
          \hat{k}^{\prime\, 2}_{\perp s}/\hat{x}^\prime_s
          \= \frac{1-\xb_1}{1-\xi^2} \vdd/2 + \frac2{1-\xb_1}
          ({\bf k}_{1\perp}+\vd/2)^2\,.
\label{inequality} \ee
Each term in this inequality is positive. Therefore, each one is
separately bounded by $2\Lambda^2$. The inequality for the second
term on the right hand side of \req{assumption-1} compels ${\bf
k}_{1\perp}\simeq -\vd/2$ for large $m_c$. This result can be
approximated by the more convenient form
\be {\bf k}_{1\perp} \simeq -x_1\vd/2  \quad {\rm for} \quad  m_c
\to \infty \ee
since the relevant values of $\xb_1$ are about $x_0$, i.e.\ large.
Analogously, one obtains ${\bf k}_{1\perp}^\prime \simeq
x_1^\prime \vd/2$. Approximating $\xb_1$ by the peak position
$x_0$, we find from the first term in \req{inequality}
\be \vdd \lsim 4\,\frac{\Lambda^2}{1-x_0}(1-\xi^2)\,. \ee
In the formal limit of $m_c\to \infty$ where $x_0$ tends to 1,
this inequality is respected for any fixed value of $\vdd$. Even
for a realistic value of about 0.6 for $x_0$ the range of allowed
values for $\vdd$ is sufficiently large for our purpose of
studying the process $p\bar{p}\to \Lambda_c^+ \bar{\Lambda}_c^-$.
Finally, the assumption of bounded virtualities, $\mid k_1^{2}\mid
\leq \Lambda^2$ and $\mid k_1^{\prime 2}-m_c^2\mid \leq
\Lambda^2$, provides $k_1^-\simeq x_1 p^-$ and an analogous result
for $k_1^{\prime\, -}$.

The above considerations lead to the following approximation of
the momenta of the active partons \req{parton-momenta}
(${k}_1^{(\prime)+}=x_1^{(\prime)+}p^{(\prime)+}$,
$\;{k}_2^{(\prime)-}=x_2^{(\prime)-} q^{(\prime)-}$):
\ba k_1 \to \bar{k}_1 &=& \left[\;{k}_1^+\,,\, \frac{x_1^2
\vdd}{8{k}_1^+}\,,
                     -\frac12 x_1\vd\right]\,, \nn\\[0.1em]
k^\prime_1 \rightarrow \bar{k}^\prime_1 &=&\left[\;{k}_1^{\prime +},
                      \frac{m_c^2+x_1^{\prime 2}\vdd/4}
                      {2{k}_1^{\prime +}}, \frac12 x_1^\prime\vd \right],\nn\\[0.1em]
k_2 \rightarrow \bar{k}_2 &=& \left[\;\frac{x_2^2\vdd}{8{k}_2^-}\,,
                    \,{k}_2^-\,, \phantom{-}\frac12 x_2\vd\right],\nn\\[0.1em]
k^\prime_2 \rightarrow \bar{k}^\prime_2
&=&\left[\frac{m_c^2+x_2^{\prime 2}\vdd/4}{2{k}_2^{\prime -}},
                                           {k}_2^{\prime -}, -\frac12 x_2^\prime\vd\right].
\label{parton-mom}
\ea
We introduce the subscripts 1 and 2 in order to distinguish the partons
from the $p\to\Lambda_c^+$ and the  $\bar{p}\to\bar{\Lambda}{}_c^-$ vertex.
It is to be emphasized that the parton momenta in this approximation are
on-shell and lie in the scattering plane formed by the baryon momenta.

%%%%%%%%%%%%%%%%%%%%%%%%%%%%%%%%%%%%%%%%%%%%%%%%%%%%%%%%%%%%%%%%%%%%%%
\subsection{Calculation of the double handbag}
%%%%%%%%%%%%%%%%%%%%%%%%%%%%%%%%%%%%%%%%%%%%%%%%%%%%%%%%%%%%%%%%%%%%%%
The amplitude for the hadron-hadron scattering process obviously
reads,
\ba {\cal M} &=& \sum_{\rm a_i^{(\prime)},\, \alpha_i^{(\prime)}}
\int d^4 k_1^{\mathrm{av}} \theta(k_1^{\mathrm{av}+})
                    \int \frac{d^4 z_1}{(2\pi)^4} e^{ik_1^{\mathrm{av}} z_1}
                    \int d^4 k_2^{\mathrm{av}} \theta(k_2^{\mathrm{av}-})
                    \int \frac{d^4 z_2}{(2\pi)^4} e^{ik_2^{\mathrm{av}} z_2} \nn\\[0.2em]
    &\times& \langle p^\prime\mu^\prime| T \bar{\Psi}^c_{a^\prime_1\alpha^\prime_1}(-z_1/2)
             \Psi^u_{a_1\alpha_1}(z_1/2)|p\mu\rangle\;
             \widetilde{H}_{a_i^{(\prime)} \alpha^{(\prime)}_i}
             (k_1^{(\prime)}, k_2^{(\prime)})\;
             \nn\\[0.2em]
    &\times&       \langle q^\prime\nu^\prime| T \bar{\Psi}^u_{a_2\alpha_2}(z_2/2)
                    \Psi^c_{a^\prime_2\alpha^\prime_2}(-z_2/2)|q\nu\rangle
                    ,
\label{ampl} \ea
where we have used $k_i^{\mathrm{av}}:=(k_i+k_i^\prime)/2$,
$(k_1-k_1^\prime)=(p-p^\prime)$, and
$(k_2-k_2^\prime)=(q-q^\prime)$. It is understood that the proton
emits a parton with momentum $k_1$, helicity $\lambda_1$ and color
$a_1$ (cf.\ Fig.\ \ref{fig:handbag}). It undergoes a hard
scattering with a corresponding antiparton ($k_2$, $\lambda_2$,
$a_2$) emitted from the antiproton. The scattered partons,
characterized by $k^\prime_1$, $\lambda^\prime_1$, $a^\prime_1$
and $k^\prime_2$, $\lambda^\prime_2$, $a^\prime_2$, are reabsorbed
by the remainders of the proton or antiproton and form the
$\Lambda_c^+$ or $\bar{\Lambda}{}_c^-$, respectively. The labels
$\alpha^{(\prime)}_i$ are spinor labels. For the ease of writing
we will omit the color and spinor labels in the following wherever
this can be done without loss of unambiguity. The active $u$ and
$\bar{u}$ quarks are treated as massless quarks. We have not to
consider an antiquark contribution in the above baryon matrix
elements as is necessary in Compton scattering \cite{DFJK1}. The
quarks in \req{subprocess} are associated with the $p\to
\Lambda_c^+$ vertex, the antiquarks with the
$\bar{p}\to\bar{\Lambda}{}_c^-$ one.

Since the parton-parton scattering is dominated by a large scale,
the heavy quark mass, we can  neglect the variation of transverse
and minus (plus) components of $k_i$ and $k^\prime_i$ in the hard
scattering kernel $\widetilde{H}$ and replace the parton momenta
by the on-shell vectors $\bar{k}_i^{(\prime)}$ \req{parton-mom}.
With this replacement of the parton momenta the integrations over
$k_1^{\mathrm{av}-}$, $k_2^{\mathrm{av}+}$ and
$\vk{}_i^{\mathrm{av}}\,, i=1,2\, ,$ can be performed explicitly
\ba &&  \int d^4 k_1^{\mathrm{av}}
\theta(k_1^{\mathrm{av}+})
                    \int \frac{d^4 z_1}{(2\pi)^4} e^{ik_1^{\mathrm{av}} z_1}
                    \int d^4 k_2^{\mathrm{av}} \theta(k_2^{\mathrm{av}-})
   \int \frac{d^4 z_2}{(2\pi)^4} e^{ik_2^{\mathrm{av}} z_2} f(\bar{k}_1,z_1,\bar{k}_2,z_2)
   \nn\\[0.2em]
&& = \int d k_1^{\mathrm{av}+} \theta(k_1^{\mathrm{av}+}) \int
\frac{d z_1^-}{2\pi} e^{ik_1^{\mathrm{av}+} z_1^-}
     \int d k_2^{\mathrm{av}-} \theta(k_2^{\mathrm{av}-})
     \int \frac{d z_2^+}{2\pi} e^{ik_2^{\mathrm{av}-} z_2^+}
         f(\bar{k}_1,\bar{z}_1,\bar{k}_2,\bar{z}_2)\,.\nn\\
\label{eq:integration}
\ea
We see that the relative distance of the fields in the matrix elements
is forced on the light-cone, $\bar{z}_1=[0,z_1^-,\tr0]$ and
$\bar{z}_2=[z_1^+,0,\tr0]$. After this the time-ordering of the fields
can be dropped \cite{diehl98}.

To show that the plus components indeed dominate the scattering we
use the fact that the proton-parton amplitudes, described by the
soft matrix elements, can be written as the amplitude for a (anti)
proton with momentum $p$ ($q$) emitting the active parton with
momentum $\bar{k}_1 (\bar{k}_2)$ and a number of on-shell
spectators times the corresponding conjugate amplitude for momenta
$p^\prime (q^\prime)$ and $\bar{k}^\prime_1 (\bar{k}^\prime_2)$,
summed over all spectator configurations; this corresponds to
inserting a complete set of intermediate states between the quark
and the antiquark fields in the matrix elements.

Using the technique of transverse boosts, we can transform to a frame
where $\bar{k}_1$ has zero transverse and minus components:
\be \bar{k}_1 \= [{k}_1^+, 0, \tr 0]\,. \ee
Using the general result
\be \frac12 (\gamma^-\gamma^+ + \gamma^+\gamma^- ) =1\,, \ee
and the energy projector for quarks~\footnote{ We use the Dirac
spinors as defined in \ci{DKV3}. The spinors refer to states with
definite light-cone helicities \ci{kogut}.}
\be \sum_{\lambda_1}
u(\bar{k}_1,\lambda_1)\bar{u}(\bar{k}_1,\lambda_1)\=
                              \bar{k}_1\cdot\gamma \= k_1^+\gamma^- \,,
\label{eq:proj}
\ee
we can write
\ba \Psi^u(z_1/2) &=& \frac12 (\gamma^-\gamma^+ +
\gamma^+\gamma^-)\Psi^u(z_1/2)
         \= \frac{1}{2{k}_1^+} \sum_{\lambda_1} \left[
    u(\bar{k}_1,\lambda_1)\left(\bar{u}(\bar{k}_1,\lambda_1)\gamma^+\Psi^u(z_1/2)\right)
        \right. \nn\\
   &+& \left. \gamma^+u(\bar{k}_1,\lambda_1)\left(\bar{u}(\bar{k}_1,\lambda_1)\Psi^u(z_1/2)
                 \right)\right] \,.
\label{eq:decomp}
\ea
Due to the central assumption of restricted parton virtualities and
transverse momenta \req{assumption-1} there are large plus components at the
proton-parton vertex but no large kinematical invariant. Because of this the
first term in \req{eq:decomp} dominates over the second one and the latter can
be neglected. Since a transverse boost leaves the plus component of any vector
unchanged, \req{eq:decomp} holds in the overall symmetric frame too.
Similar considerations for the outgoing partons run into complications with
the quark mass. A transverse boost to a frame where $\bar{k}^\prime_1$ has no
transverse component, leads to
\be
 \bar{k}^\prime_1 \longrightarrow \left[{k}^{\prime +}_1,
\frac{m_c^2}{2{k}_1^{\prime +}},\tr0\right]\,, \ee
with a non-vanishing minus component due to the finite charm-quark
mass. The energy projector reads
\be \sum_{\lambda^\prime_1} u(\bar{k}^\prime_1,\lambda^\prime_1)
                     \bar{u}(\bar{k}^\prime_1,\lambda^\prime_1)\=
                              \bar{k}^\prime_1\cdot\gamma + m_c \=
{k}_1^{\prime +}\gamma^- + \frac{m_c^2}{2{k}_1^{\prime +}}\gamma^+ + m_c
\label{eq:proj-mass}
\ee
in that frame. For the charm field we can therefore write (with $(\gamma^+)^2=0$)
\ba \Psi^c(-z_1/2) &=& \frac{1}{2{k}_1^{\prime +}}
\sum_{\lambda^\prime_1} \left [
       u(\bar{k}^\prime_1,\lambda^\prime_1)
       \left(\bar{u}(\bar{k}^\prime_1,\lambda^\prime_1)\gamma^+\Psi^c(-z_1/2)\right)
                                \right.\nn\\
       &+& \left.\gamma^+\Big\{ u(\bar{k}^\prime_1,\lambda^\prime)
        (\bar{u}(\bar{k}^\prime_1,\lambda^\prime_1)\Psi^c_{a^\prime_1}(-z_1/2))
                                            -2m_c\Psi^c(-z_1/2)\Big\}\right]\,.
\label{eq:decomp-mass}
\ea
Using the plane-wave decomposition of the charm-quark field it can
be shown that the term within curly brackets vanishes identically
if it is applied to a $\Lambda_c^+$ Fock state in which the charm
quark carries momentum $\bar{k}^\prime_1$. But these are just the
states we need in the sequel to describe the absorption of the
charm quark, that leaves the hard scattering process with momentum
$\bar{k}^\prime_1$, by the $\Lambda_c^+$. We therefore conclude
again that the first term in (\ref{eq:proj-mass}) dominates over
the one within square brackets so that the latter can be
neglected.
%Assuming for the
%moment that this term is indeed small in the frame where
%$\bar{k}^\prime_1$ has no transverse component.
In the overall symmetric frame this term is small as well and  we
have
\be
\Psi^c(-z_1/2) \= \frac{1}{2{k}_1^{\prime +}} \sum_{\lambda^\prime_1} %\left [
                        u(\bar{k}^\prime_1,\lambda^\prime_1)
   \left(\bar{u}(\bar{k}^\prime_1,\lambda^\prime_1)\gamma^+\Psi^c(-z_1/2)\right)\,.
\label{eq:decomp-c}
\ee

Employing the light-cone spinors we can write for the helicity non-flip and
flip configuration at the $p\to \Lambda_c^+$ vertex in the  symmetric frame
\ba \frac{\bar{u}(\bar{k}^\prime_1,\lambda_1) \gamma^+
    u(\bar{k}_1,\lambda_1)}{2\sqrt{{k}^+_1 k_1^{\prime +} }} &=& 1\,, \nn\\
%\frac{\bar{v}(\bar{k}_2,\lambda_2) \gamma^-
%    v(\bar{k}^\prime_2,\lambda_2)}{2{k}^{\prime -}_2} &=& - 1\,,
\frac{\bar{u}(\bar{k}^\prime_1,-\lambda_1)\imath \sigma^{+j}
    u(\bar{k}_1,\lambda_1)}{2\sqrt{k^+_1 k_1^{\prime +}}} &=&
                          \imath(2\lambda_1\imath)^j\,,
%\frac{\bar{v}(\bar{k}_2,\lambda_2)\imath \sigma^{-j}
%    v(\bar{k}^\prime_2,-\lambda_2)}{2\bar{k}^{\prime -}_1} &=&
%                         -\imath(2\lambda_2\imath)^j \,,
\label{eq:insertions} \ea
where $j=1,2$ and
\be \sigma^{\pm j} \= \imath\gamma^{\pm}\gamma^j\,. \ee
Of course, analogous expressions hold for the antiparticle spinors
(with an additional minus sign). The helicity projector for the
light quarks reads
\be u(\bar{k}_1,\lambda_1)\bar{u}(\bar{k}_1,\lambda_1)
                          \= \bar{k}_1\cdot \gamma (1-2\lambda_1\gamma_5)/2\,.
\ee
On the other hand, for the charm quark we have
\be
u(\bar{k}^\prime_1,\lambda^\prime_1)\bar{u}(\bar{k}^\prime_1,\lambda^\prime_1)\=
         \Big[\bar{k}^\prime_1\cdot \gamma + m_c\Big]
              (1+\gamma_5 S_1\cdot \gamma)/2\,,
\label{eq:spin-projector}
\ee
where the covariant spin vector $S_1$ reads
\be S_1\=\frac{2\lambda^\prime_1}{m_c}\,\big[\bar{k}^\prime_1 -
\frac{m_c^2}{{k}_1^{\prime +}}\,n^-\big]\,. \ee
Here, $n^-\=[0,1,\tr{0}]$ is a unit vector. In the limit $m_c\to
0$ the spin projector \req{eq:spin-projector} turns into the usual
helicity projector.

Now we can simplify the products of quark fields occurring in
\req{ampl}. In fact, forming the product of the first terms in
\req{eq:decomp} and \req{eq:decomp-mass} and inserting
\req{eq:insertions} between the two terms involving $\gamma^+$,
one obtains after a little algebra
\ba
\bar{\Psi}^c_{\alpha^\prime_1}(-z_1/2)\,\Psi^u_{\alpha_1}(z_1/2)
&=&
                      \frac1{4k^+_1k^{\prime +}_1}
        \sum_{\lambda_1\lambda^\prime_1} \Big(\bar{\Psi}^c(-z_1/2)\,\gamma^+
                       u(\bar{k}^\prime_1,\lambda^\prime_1)\Big)  \nn\\
 &\times&                \Big(\bar{u}(\bar{k}_1,\lambda_1)\gamma^+\Psi^u(z_1/2)\Big)
        \bar{u}_{\alpha^\prime_1}(\bar{k}^\prime_1,\lambda^\prime_1)
          u_{\alpha_1}(\bar{k}_1,\lambda_1)  \nn\\
     \longrightarrow \hspace*{0.15\tw} && \hspace*{-0.20\tw}
     \frac1{2\sqrt{k^+_1 k^{\prime +}_1}}
   \sum_{\lambda_1}\left\{ \bar{\Psi}^c(-z_1/2)\,\gamma^+ \frac{1+2\lambda_1\gamma_5}{2}
                 \Psi^u(z_1/2) \bar{u}_{\alpha^\prime_1}(\bar{k}^\prime_1,\lambda_1)
           u_{\alpha_1}(\bar{k}_1,\lambda_1)\right.\nn\\
      && \hspace*{-0.20\tw}   -\left.\imath(-2\lambda_1\imath)^j
           \bar{\Psi}^c(-z_1/2)\,\imath\sigma^{+j}
                   \frac{1+2\lambda_1\gamma_5}{2}
                 \Psi^u(z_1/2) \bar{u}_{\alpha^\prime_1}(\bar{k}^\prime_1,-\lambda_1)
           u_{\alpha_1}(\bar{k}_1,\lambda_1)\right\}\,.\hspace*{0.08\tw}
\label{eq:replace-up}
\ea
The quark-mass and the transverse-momentum terms drop out since they
always come with products of two $\gamma^+$ matrices. Analogously one
finds for the $\bar{p}\to\bar{\Lambda}{}_c^-$ vertex
\ba
 \bar{\Psi}^u_{\alpha_2}(z_2/2)\,\Psi^c_{\alpha^\prime_2}(-z_2/2) &=&
             -\frac1{2\sqrt{k^-_2 k^{\prime -}_2}}
    \sum_{\lambda_2}  \nn\\
     && \hspace*{-0.2\tw} \times\left\{
   \bar{\Psi}^u(z_2/2)\,\gamma^- \frac{1-2\lambda_2\gamma_5}{2}
                 \Psi^c(-z_2/2) \bar{v}_{\alpha_2}(\bar{k}_2,\lambda_2)
           v_{\alpha^\prime_2}(\bar{k}^\prime_2,\lambda_2)\right.\nn\\
   && \hspace*{-0.2\tw}  -\left.\imath(-2\lambda_2\imath)^j
           \bar{\Psi}^u(z_2/2)\,\imath\sigma^{-j}
                   \frac{1+2\lambda_2\gamma_5}{2}
                 \Psi^c(-z_2/2) \bar{v}_{\alpha_2}(\bar{k}_2,\lambda_2)
           v_{\alpha^\prime_2}(\bar{k}^\prime_2,-\lambda_2)\right\} .
\label{eq:replace-low} \ea
Combining now \req{ampl}, \req{eq:integration},
\req{eq:replace-up} and \req{eq:replace-low} and using
$\lambda_2=-\lambda_1$, a property of the subprocess amplitude
which follows from the fact that massless quarks do not flip their
helicities in interactions with photons and/or gluons, one arrives
at
\vfill\break
\ba {\cal M} &=& -
   \int d k_1^{\mathrm{av}+} \theta(k_1^{\mathrm{av}+}) \int \frac{d z_1^-}{2\pi}
   e^{ik_1^{\mathrm{av}+} z_1^-}
     \int d k_2^{\mathrm{av}-} \theta(k_2^{\mathrm{av}-})
     \int \frac{d z_2^+}{2\pi} e^{ik_2^{\mathrm{av}-} z_2^+}
           \frac1{4\sqrt{k_1^+k_1^{\prime +} k_2^- k_2^{\prime -}}}  \nn\\[0.2em]
    &\times&  \sum_{\lambda_1}\left\{
\langle p^\prime\mu^\prime\mid \bar{\Psi}^c(-\bar{z}_1/2) \gamma^+ \frac{1+2\lambda_1\gamma_5}{2}
             \Psi^u(\bar{z}_1/2)\mid p \mu\rangle\, \right.\nn\\
     &\times& \left.   \langle q^\prime\nu^\prime\mid \bar{\Psi}^u(\bar{z}_2/2) \gamma^-
                     \frac{1+2\lambda_1\gamma_5}{2}
             \Psi^c(-\bar{z}_2/2)\mid q \nu\rangle\,
             H_{\lambda_1 -\lambda_1, \lambda_1-\lambda_1}(\bar{k}_1,\bar{k}_2)\right.   \nn\\[0.2em]
&-& \left.
               \langle p^\prime\mu^\prime\mid \bar{\Psi}^c(-\bar{z}_1/2) \imath\sigma^{+j}
         \frac{1+2\lambda_1\gamma_5}{2}
          \Psi^u(\bar{z}_1/2)\mid p \mu\rangle\,  \right.\nn\\[0.2em]
&\times& \left. \langle q^\prime\nu^\prime\mid \bar{\Psi}^u(\bar{z}_2/2) \imath\sigma^{-j}
                      \frac{1-2\lambda_1\gamma_5}{2}
             \Psi^c(-\bar{z}_2/2)\mid q \nu\rangle\,
       H_{-\lambda_1 \lambda_1, \lambda_1 -\lambda_1}(\bar{k}_1,\bar{k}_2)
                                           \right.\nn\\[0.2em]
  &-&\left.\imath (2\lambda_1\imath)^j
\langle p^\prime\mu^\prime\mid \bar{\Psi}^c(-\bar{z}_1/2) \gamma^+ \frac{1+2\lambda_1\gamma_5}{2}
             \Psi^u(\bar{z}_1/2)\mid p \mu\rangle\, \right.\nn\\[0.2em]
&\times& \left.
\langle q^\prime\nu^\prime\mid \bar{\Psi}^u(\bar{z}_2/2) \sigma^{-j}
              \frac{1-2\lambda_1\gamma_5}{2} \Psi^c(-\bar{z}_2/2)\mid q
          \nu\rangle\,
          H_{\lambda_1 \lambda_1, \lambda_1-\lambda_1}(\bar{k}_1,\bar{k}_2)
                                                     \right.\nn\\[0.2em]
 &-&\left.\imath (-2\lambda_1\imath)^j
\langle p^\prime\mu^\prime\mid \bar{\Psi}^c(-\bar{z}_1/2) \sigma^{+j}
                       \frac{1+2\lambda_1\gamma_5}{2}
                  \Psi^u(\bar{z}_1/2)\mid p\mu\rangle\, \right.\nn\\[0.2em]
&\times& \left.
\langle q^\prime\nu^\prime\mid \bar{\Psi}^u(\bar{z}_2/2) \gamma^-
                       \frac{1+2\lambda_1\gamma_5}{2}
             \Psi^c(-\bar{z}_2/2)\mid q \nu\rangle\,
        H_{-\lambda_1 -\lambda_1, \lambda_1-\lambda_1}(\bar{k}_1,\bar{k}_2)
                                    \right\}\,.
\label{eq:A}
\ea
The subprocess amplitudes are defined by
\be H_{\lambda_1^\prime \lambda_2^\prime ,
\lambda_1\lambda_2}(\bar{k_1},\bar{k_2}) \=
 \bar{v}(\bar{k}_2,\lambda_2)\bar{u}(\bar{k}^\prime_1,\lambda^\prime_1)
\widetilde{H}(\bar{k}_1^{(\prime)},\bar{k}_2^{(\prime)})
v(\bar{k}^\prime_2,\lambda_2^\prime)u(\bar{k}_1,\lambda_1)\,,
\label{def:subprocess-amplitudes} \ee
where the hard scattering kernel
$\widetilde{H}(\bar{k}_1^{(\prime)},\bar{k}_2^{(\prime)})$ has to
be evaluated for  parton momenta as given in (\ref{parton-mom}).

%%%%%%%%%%%%%%%%%%%%%%%%%%%%%%%%%%%%%%%%%%%%%%%%%%%%%%%%%%%%%%%%%%%%%%%%%
\subsection{Peaking approximation, GPDs and form factors}
\label{sec:peaking}
%%%%%%%%%%%%%%%%%%%%%%%%%%%%%%%%%%%%%%%%%%%%%%%%%%%%%%%%%%%%%%%%%%%%%%%%%
According to our discussion in Sect.2.1 we set
$k^{(\prime)+}_1=x_1^{(\prime)}p^{(\prime)+}$,
$k^{(\prime)-}_2=x_2^{(\prime)}q^{(\prime)-}$. In order to produce
a $\cbc$ pair the subprocess Mandelstam variable, $\hat
s\=x_1x_2s$, must be larger than $4m_c^2$. For the sake of the
argument consider a symmetric configuration of $p\to \Lambda_c^+$
and $\bar{p}\to \bar{\Lambda}{}_c^-$ which, with regard to the
expected behavior of the GPDs, namely a marked narrow peak
centered around $x_0$, is an appropriate assumption. We then have
the kinematical requirement
\be
    x \geq 2x_0 \frac{M}{\sqrt{s}}\,.
\label{eq:kinematic-requirement} \ee
As the comparison with \req{skew} in the appendix reveals, the
inequality \req{eq:kinematic-requirement} requires $x$ to be
larger than the skewness in the forward hemisphere. With the help
of \req{individual-fractions} one sees that the inequality
\req{eq:kinematic-requirement} also implies $\xb\geq 2(1+\xi)x_0
M/\sqrt{s}-\xi$ and, hence, $\xb\geq \xi$. Consequently, the ERBL
region $\xb\leq \xi$ does not contribute to the process
$p\bar{p}\to \Lambda_c^+\bar{\Lambda}_c^-$. For large $s$ the
lower bound of $\xb$ becomes $\xb \geq 2 x_0 M/\sqrt{s}$, i.e.\ it
tends to zero for $s\to\infty$ as does $\xi$, see Eq.\ \req{skew}.

Given the expected shape of the GPDs it is clear that only regions
close to the peak position contribute to any degree of
significance. For the hard partonic subprocess we therefore employ
the peaking approximation, i.e.\ for the momenta \req{parton-mom}
we use $x_1\simeq x_2\simeq x_0$. From the parton momenta
\req{parton-mom} it then follows  that the Mandelstam variables
for the subprocess are proportional to those of the full process
\be \hat{s} \simeq x_0^2 s\,,\qquad \hat{u} \simeq x_0^2u\,,
\qquad \hat{t} \simeq x_0^2\,t\,. \label{eq:pa} \ee
The sum of the three subprocess Mandelstam variables is
\be \hat{s}+\hat{u}+\hat{t} \simeq 2m_c^2\,. \label{eq:mom-con}
\ee
Thus, momentum conservation for the parton kinematics holds. The
use of the peaking approximation is quite appealing since it
nicely matches the subprocess kinematics to the one for the
hadronic process up to corrections of order $m^2/s$ and $\xi
(1-x_0)$. The accuracy of the peaking approximation increases with
increasing $s$ at fixed mass of the heavy quark and with
increasing $m_c$ for fixed values of $m_c^2/s$. For the physical
charm system the peaking approximation holds, for instance, for
\req{eq:mom-con} or the cosine of the c.m.s.\ scattering angle
$\theta$ with an accuracy better than $10 (20)\%$ for
$s=30(25)\,\gev^2$ in the forward hemisphere. This means that
$\cos{\theta}$ either evaluated from $s$ and $t$ or from $\sh$ and
$\th$ only differs by less than the quoted accuracy. The use of
the peaking approximation therefore implies that our approach can
only be applied to values of $s$ well above the kinematical
threshold. The natural requirement that a substantial part of the
peak region of the GPDs contributes to the form factors
necessitates $\xi\ll x_0$ which in turn implies not too small
values of $s$. Numerical checks reveal that this requirement is
satisfied provided $s\gsim 23\,\gev^2$.

The Fourier transforms of the soft hadronic matrix
elements~\footnote{ In order to have  manifestly charge
conjugation symmetric GPDs (see App.\ \ref{app-B}) we have to add
the corresponding antiparticle contribution to the  bilocal matrix
elements of the field operators in \req{eq:A} (see definitions
\req{def:V} and \req{V-ME} and analogous ones for the other GPDs)
which can be done since these contributions are zero because of
the neglect of intrinsic charm.} occurring in \req{eq:A} define
the GPDs as we discuss in App.\ \ref{app-B} \req{V-ME}, \req{A-ME}
and \req{eq:GPD-flip}. Using
\be \int
dk_1^{\mathrm{av}+}\frac{\theta(k_1^{\mathrm{av}+})}{\sqrt{k_1^+k_1^{\prime
+}}} \=
 \int_\xi^1\frac{d\xb_1}{\sqrt{\xb_1^2-\xi^2}}\,,
\label{eq:inte} \ee
and  an analogous expression for the integral over
$k_2^{\mathrm{av}-}$ as well as the covariant decompositions
\req{V-FF}, \req{A-FF} and  \req{eq:GPD-flip}, one sees that the
integrations lead to $1/\xb_1$ moments of the GPDs. For instance,
for the vector matrix element we arrive at
\be \int_\xi^1 \frac{d\xb_1}{\sqrt{\xb_1^2-\xi^2}} {\cal
H}_{\mu'\mu}^{\,cu}
                \= R_V(\xi,t) \bar{u}(p',\mu') \gamma^+ u(p,\mu)
             + R_T(\xi,t) \bar{u}(p',\mu')\frac{\imath\sigma^{+\nu}\Delta_\nu}{M+m}
         u(p,\mu)\, .
 \label{eq:ann-FF}
\ee
The form factors in \req{eq:ann-FF} are defined by
\be
         R_i(\xi,t) \= \int_\xi^1 \frac{d\xb_1}{\sqrt{\xb_1^2-\xi^2}} F_i(\xb_1,\xi,t)\,,
\label{eq:annFF} \ee
where $F_i$ and $R_i$ are generic for any of the $u\to c$ or
$\bar{u} \to \bar{c}$ transition GPDs
$H,E,\widetilde{H},\widetilde{E}, H_T,$ $\cdots$, and the
associated form factors $R_V, R_T, R_A, R_P, S_T, \cdots$. We
apply here the notation proposed in \ci{passek04}. Due to the
kinematical requirement $\xb\geq\xi$ the lower limit of the
integration is taken to be $\xi$ instead of zero. This leads to a
mild $\xi$ (or $s$) dependence of the form factors. For
$s\to\infty$ the $\xi$ dependence of the form factors disappears
since in this limit $\xi$ tends to zero like $(M^2-m^2)/s$, as can
be seen from \req{skew}.  The form factors $R_i$ stand for linear
combinations of more commonly used form factors, as can be seen
from Eqs. \req{vector-sum-rules} and \req{axial-sum-rules} in the
Appendix.
%We do not need these form factors separately, only these
Only such linear combinations occur in the process amplitudes.
Due to the time reversal invariance the transition form factors
are real valued.

It may seem plausible to choose the peaking approximation for the factor
$1/\sqrt{\xb_i^2 -\xi^2}$ in Eqs.\ \req{eq:A} and \req{eq:inte} too and to use
for the form factors the expression
\be
  R_i \simeq \frac1{\sqrt{x_0^2-\xi^2}}\, \int_\xi^1 d\xb
  F_i(\xb,\xi,t)\, .
\label{eq:FFappr}
\ee
For GPDs concentrated in a narrow range of large $\xb$, centered
around $x_0$, the difference between \req{eq:annFF} and
\req{eq:FFappr} is expected to be small. We will return to this
issue when we present our model GPDs.

Working out the various hadronic covariants occurring in \req{eq:ann-FF} and the
other baryon matrix elements in the symmetric frame, we find the following
results for quark helicity non-flip (conventions - helicities of the outgoing
(anti)baryons appear always first; $\xb$ is either $\xb_1$ or $\xb_2$)
\ba \int_\xi^1\frac{d\xb}{\sqrt{\xb^2-\xi^2}}\;{\cal
H}^{cu}_{\mu\mu} \;(\,{\cal H}^{\overline{cu}}_{\mu\mu}\,) &=&
\delta 2 \bar{p}^+\sqrt{1-\xi^2}\,R_{V{\rm eff}}\,,\nn\\[0.2em]
\int_\xi^1\frac{d\xb}{\sqrt{\xb^2-\xi^2}}\,{\cal H}^{cu}_{-\mu\mu}
\,({\cal H}^{\overline{cu}}_{-\mu\mu}) &=&
    \delta\varepsilon \frac{2\bar{p}^+}{\sqrt{1-\xi^2}}\,
                         \frac{\Delta_\perp}{M+m}\,R_T\,,\nn\\[0.2em]
\int_\xi^1\frac{d\xb}{\sqrt{\xb^2-\xi^2}}\;\widetilde{{\cal H}}^{cu}_{\mu\mu}
\;(\,\widetilde{{\cal H}}^{\overline{cu}}_{\mu\mu}\,) &=&
\varepsilon\,2\bar{p}^+\sqrt{1-\xi^2}\,
    \Big[R_A-\frac{\xi}{1-\xi^2}\widetilde{M} R_P\Big]\,,\nn\\[0.2em]
\int_\xi^1\frac{d\xb}{\sqrt{\xb^2-\xi^2}}\,\widetilde{{\cal H}}^{cu}_{-\mu\mu}
\,(\widetilde{{\cal H}}^{\overline{cu}}_{-\mu\mu}) &=&
    2\bar{p}^+\,\frac{\xi}{\sqrt{1-\xi^2}}\,\frac{\Delta_\perp}{M+m}\,R_P\,,
\label{eq:vertex-nf}
\ea
\be \widetilde{M}\=\frac{(1+\xi)M-(1-\xi)m}{M+m}\,,\qquad R_{V{\rm
eff}}\= R_V- R_T\frac{\xi}{1-\xi^2}\widetilde{M}\,. \ee
Here, $\varepsilon=(-1)^{\mu-1/2}$, $\delta=(-)1$ for (anti)
baryons.

For quark helicity flip we define the combinations ($\lambda_i$ refers to the
helicity of the emitted parton)
\ba {\cal H}^{Tcu}_{\lambda_1\mu^\prime\mu}&=&\frac12\Big[{\cal
H}^{Tcu}_{1\mu^\prime\mu}
     -2\lambda_1\imath{\cal H}^{Tcu}_{2\mu^\prime\mu}\Big]\,, \nn\\
{\cal H}^{T\overline{c u}}_{\lambda_1\nu^\prime\nu}
     &=&\frac12\Big[{\cal H}^{T\overline{c u}}_{1\nu^\prime\nu}
     -2\lambda_1\imath{\cal H}^{T\overline{c u}}_{2\nu^\prime\nu}\Big]\,,
\ea
of helicity-flip GPDs \ci{diehl01} which are defined in
\req{eq:GPD-flip} and \req{eq:GPD-flip-anti}. These combinations
make the helicity projection explicit since they correspond to the
Dirac structure $\imath/2\sigma^{\pm 1}(1\pm2\lambda_1\gamma_5)$
(see \req{ab-a5}). Their moments read
\ba \int_\xi^1\frac{d\xb}{\sqrt{\xb^2-\xi^2}}\,{\cal
H}^{Tcu}_{\lambda_1\mu\mu} \,({\cal
H}^{T\overline{cu}}_{\lambda_1\mu\mu}) &=&
     \delta\frac{\bar{p}^+}{\sqrt{1-\xi^2}}\frac{\Delta_\perp}{M+m}\,\Big\{
         \,(M+m)\frac{(1+\xi)M+(1-\xi)m}{2mM}\,S_S \nn\\
&+& \big[1+2\lambda_1\delta\varepsilon\xi\big]\, S_{V_1}
         -\big[\xi+2\lambda_1\delta\varepsilon\big]\,S_{V_2}
              \Big\} \, ,    \nn\\[0.2em]
\int_\xi^1\frac{d\xb}{\sqrt{\xb^2-\xi^2}}\,{\cal
H}^{Tcu}_{\lambda_1-\mu\mu} \,({\cal
H}^{T\overline{cu}}_{\lambda_1-\mu\mu}) &=&
    -\frac{\bar{p}^+}{\sqrt{1-\xi^2}}\,\Big\{(1-\xi^2)(\delta\varepsilon+2\lambda_1)\,S_T
  + \delta\varepsilon \frac{\vdd}{mM}\,S_S\nn\\
  &-& (\delta\varepsilon+2\lambda_1)\, \widetilde{M}\,
      \Big[\xi\, S_{V_1} - S_{V_2}\Big]\phantom{\frac{\vdd}{mM}}\Big\}\,,
\label{eq:vertex-f} \ea
and represent the quark helicity-flip matrix elements occurring in
\req{eq:A}. The form factors are defined in \req{eq:annFF}.

The matrix elements \req{eq:vertex-nf} and \req{eq:vertex-f} vanish for
$t\to t_0$ at least as
\be
     \propto \sqrt{t_0-t}^{\;\mid
     \lambda'_1-\mu'-\lambda_1+\mu\mid}\,.
\label{eq:o-a-m}
\ee
This behavior reflects the fact that the mismatch of the quark and
hadron helicities in these matrix elements is to be compensated by
a corresponding number of units of orbital angular momenta in
order to ensure angular momentum conservation. Closer inspection
of the matrix elements for the various helicity configurations
reveals that only $R_V$, $R_A$ and $S_T$ (corresponding to the
GPDs $H$, $\widetilde{H}$ and $H_T$) do not involve orbital
angular momentum and therefore do not vanish for $t\to t_0$. The
form factors $R_T$, $R_P$, $S_{V1}$ and $S_{V2}$ require one unit
of orbital angular momentum while $S_S$ (corresponding to the GPD
$\widetilde{H}_T$) requires even two.

%%%%%%%%%%%%%%%%%%%%%%%%%%%%%%%%%%%%%%%%%%%%%%%%%%%%%%%%%%%%%%%%%%%%%%%%%%
\section{Amplitudes}
\label{sec:amplitudes}
%%%%%%%%%%%%%%%%%%%%%%%%%%%%%%%%%%%%%%%%%%%%%%%%%%%%%%%%%%%%%%%%%%%%%%%%%%
\subsection{Subprocess amplitudes}
\label{sec:subprocess}
%%%%%%%%%%%%%%%%%%%%%%%%%%%%%%%%%%%%%%%%%%%%%%%%%%%%%%%%%%%%%%%%%%%%%%%%%%
The amplitudes $H_{\lambda'_1\lambda'_2,\lambda_1\lambda_2}$ for the subprocess
\req{subprocess} calculated to lowest order of perturbative QCD, read
\ba H_{+-,+-} &=&-\frac{8\pi\als(\sh)}{\sh}\,
         \frac{\uh_1-\uh}{\sqrt{1-4m_c^2/\sh}}\,,\nn\\[0.2em]
H_{-+,+-} &=& \frac{8\pi\als(\sh)}{\sh}\,\frac{\th_0-\th}
              {\sqrt{1-4m_c^2/\sh}}\,,\nn\\[0.2em]
H_{++,+-} &=& \frac{16\pi\als(\sh)}{\sh}\,\frac{m_c}{\sqrt{\sh}}
           \sqrt{\frac{(\th_0-\th)(\uh_1-\uh)}{1-4m_c^2/\sh}}\,,\nn\\[0.2em]
H_{++,-+} &=& H_{++,+-}\,.
\label{subprocess-amp}
\ea
Since we consider the $u$ quark as a massless particle all amplitudes
with the same helicities of the $u$ and $\bar{u}$ quark are zero.
Helicity amplitudes other than those given in \req{subprocess-amp} are
obtained by applying the relation
\be H_{-\lambda'_1 -\lambda'_2, -\lambda_1 \lambda_1} \=
    -(-1)^{\lambda'_1-\lambda'_2}\,
   H_{\lambda'_1 \lambda'_2, \lambda_1 -\lambda_1}\,,
\ee
which follows from parity invariance. With the help of the peaking
approximation we can express the subprocess amplitudes  in terms
of the kinematical variables of the full process
\ba
H_{+-,+-} &=&-8\pi\als(x_0^2s)\,\cos^2{\theta/2}\,, \nn\\
H_{-+,+-} &=&\phantom{-}8\pi\als(x_0^2s)\,\sin^2{\theta/2}\,,\nn\\
H_{++,+-} &=&\phantom{-}8\pi\als(x_0^2s)\,\frac{M}{\sqrt{s}}\sin{\theta}\,.
\ea
The quality of the peaking approximation has been discussed in Sect.\ \ref{sec:peaking}.
The QCD coupling constant $\als$ is evaluated for four flavors with
$\Lambda_{\rm QCD}=0.24\,\gev$.

The cross section for the subprocess reads
\be \frac{d\hat{\sigma}}{d\hat{\Omega}}\= \frac1{64\pi^2}
       \frac{\Lambda_{m_c}}{\hat{s}}\,\hat{\sigma}_0\,,
\ee
where $\Lambda_{m_c}$ is defined in (\ref{eq:lambdami}) and
\be \hat{\sigma}_0\=\frac14\sum_{\lambda_1^\prime
\lambda_2^\prime,\lambda_1}
      \mid H_{\lambda_1^\prime \lambda_2^\prime,\lambda_1-\lambda_1} \mid^2
       \=(4\pi\als)^2\Big[1+\cos^2{\theta} +\frac{4M^2}{s}\sin^2{\theta}\Big]\,.
\ee

To lowest order of QCD the subprocess amplitudes are real. As a
consequence any transversal polarization of the involved quarks is
zero since it is given by the imaginary part of a flip-non-flip
interference. To higher order of QCD this will change. The
subprocess amplitudes become complex \ci{goldstein} and a
$c$-quark polarization of order $\als m_c/\sqrt{\hat{s}}$ is
obtained. The polarization of the $u$ quark is zero to all orders
since it is treated as massless.

%%%%%%%%%%%%%%%%%%%%%%%%%%%%%%%%%%%%%%%%%%%%%%%%%%%%%%%%%%%%%%%%%%%%%%%%%%
\subsection{The hadronic amplitudes}
%%%%%%%%%%%%%%%%%%%%%%%%%%%%%%%%%%%%%%%%%%%%%%%%%%%%%%%%%%%%%%%%%%%%%%%%%
Inserting the definitions \req{V-ME}, \req{A-ME}, \req{eq:GPD-flip} as well as
\req{eq:GPD-flip-anti} into \req{eq:A}, we obtain for the helicity amplitudes
of the process $p\bar{p}\to \Lambda_c^+ \bar{\Lambda}_c^-$
\ba M_{\mu'\nu',\mu\nu} &=& -\left(\frac{C_F}{N_C}\right)
\,\frac1{8(\bar{p}^+)^2}\,
\int\frac{d\xb_1}{\sqrt{\xb_1^2-\xi^2}}\,\frac{d\xb_2}{\sqrt{\xb_2^2-\xi^2}}\,\Big\{  \nn\\[0.2em]
 && +H_{+-+-}\,\Big[{\cal H}^{\,cu}_{\mu'\mu}{\cal H}^{\,\overline{cu}}_{\nu'\nu}
  + \widetilde{\cal H}^{\,cu}_{\mu'\mu}\widetilde{{\cal H}}^{\,\overline{cu}}_{\nu'\nu}
        \Big]  \nn\\[0.2em]
 && -2\,H_{-++-}\,\Big[{\cal H}^{\,Tcu}_{+\mu'\mu}{\cal H}^{\,T\overline{cu}}_{+\nu'\nu}
           + {\cal H}^{\,Tcu}_{-\mu'\mu}{\cal H}^{\,T\overline{cu}}_{-\nu'\nu}
               \Big] \nn\\[0.2em]
&&  +\,  H_{+++-}\,\Big[{\cal H}^{\,cu}_{\mu'\mu}
        ({\cal H}^{\,T\overline{cu}}_{+\nu'\nu} + {\cal H}^{\,T\overline{cu}}_{-\nu'\nu})
        + \widetilde{{\cal H}}^{\,cu}_{\mu'\mu}
({\cal H}^{\,T\bar{c}\bar{u}}_{+\nu'\nu} - {\cal H}^{\,T\overline{cu}}_{-\nu'\nu})\Big]
                   \nn\\[0.2em]
&&  +\, H_{++-+}\,\Big[({\cal H}^{\,Tcu}_{+\mu'\mu}+{\cal H}^{\,Tcu}_{-\mu'\mu})
    {\cal H}^{\,\overline{cu}}_{\nu'\nu}
  +   ({\cal H}^{\,Tcu}_{+\mu'\mu}-{\cal H}^{\,Tcu}_{-\mu'\mu})
     \widetilde{{\cal H}}^{\,\overline{cu}}_{\nu'\nu}\Big]\Big\}\,.
\label{eq:combi} \ea
Here, the ratio $C_F/N_c$ is a color factor ($=4/9$). In
principle, one may express the amplitudes in terms of the form
factors using \req{eq:vertex-nf} and \req{eq:vertex-f}. However,
this will lead to a lengthy expression in terms of the eight form
factors. We refrain from doing this. The following hierarchy of
these form factors is to be expected
\be \mid R_V \mid, \mid R_A \mid, \mid S_T\mid \gg \mid R_T
\mid\,, \mid R_P \mid, \mid S_{V1}\mid, \mid S_{V2}\mid \,.
\label{eq:hierarchy} \ee
This expectation is based on the fact that the supposedly
suppressed form factors  involve orbital angular momentum of the
quarks making up the baryons. It is also supported by the model
calculations for the form factors which we will present in Sect.\
\ref{sec:GPD}.  The analogous hierarchy has also been found for
the form factors parameterizing the elastic $p\to p$ matrix
elements which, for instance, occur in wide-angle Compton
scattering \ci{DFJK1}. The reason for it is that according to
theoretical expectations and phenomenological experience, the
valence Fock states of many hadrons are dominated by parton
configurations with zero orbital angular momentum. Given the
hierarchy \req{eq:hierarchy} and the smallness of skewness
\req{skew}, we may neglect all terms $\propto \xi$ in
\req{eq:vertex-nf} and \req{eq:vertex-f}. This way the form factor
$R_P$ decouples completely. For quark helicity flip we have
additional suppression from the subprocess amplitudes
\req{subprocess-amp} in the forward hemisphere. As can be seen
from \req{eq:combi} configurations with quark helicity flip at
both hadronic matrix elements come with the amplitude $H_{-+,+-}$
and are therefore suppressed by an additional factor $\th_0-\th$.
Interference terms between flip and non-flip matrix elements are
proportional to $H_{++,\pm\mp}$ and are suppressed by
$\sqrt{\th_0-\th}\,m_c/\sh$. Due to these properties one may
neglect all form factors related to the helicity-flip GPDs with
the exception of $S_T$. The model GPDs and form factors which we
will present in Sect.~\ref{sec:GPD} support this assumption.

For later use we provide the explicit  expressions for the
helicity amplitudes by taking into account only the most important
form factors $R_V, R_A, S_T$. Inserting \req{eq:vertex-nf} and
\req{eq:vertex-f} into \req{eq:combi} we obtain the following set
of helicity amplitudes for  $p\bar{p}\to
\Lambda_c^+\bar{\Lambda}{}_c^-$
\ba M_{+\pm,+\pm} &=&
\phantom{-}\frac{C_F}{2N_c}\,(1-\xi^2)\,H_{+-,+-}\,
                  \Big[R^{\,2}_{V}\mp R_A^2\,\big]\,,\nn\\
M_{--,++} &=&\phantom{-}  0\,,\nn\\
M_{-+,+-} &=&\phantom{-} \frac{C_F}{N_c}\,(1-\xi^2) H_{-+,+-}\, S_T^2\,,\nn\\
M_{++,\pm\mp} &=& \phantom{-}\frac{C_F}{2N_c}\,(1-\xi^2)\,H_{++,\pm\mp}\,
                        S_T \Big[R_{V}+R_A\Big]\,,\nn\\
M_{\mp\pm,++} &=&  -\frac{C_F}{2N_c}\,(1-\xi^2)\,
                    H_{++,\mp\pm}\, S_T\Big[R_{V}-R_A\Big]\,.
\label{eq:lc-amplitudes} \ea
The remaining amplitudes are fixed by parity invariance
\be M_{-\mu'-\nu',-\mu-\nu} \= (-1)^{\mu'-\nu'-\mu+\nu}\,
M_{\mu'\nu',\mu\nu}\,. \ee
One can easily convince oneself that the combination $R_V+R_A$
parameterizes the soft hadronic matrix element for the situation
where the proton with a given helicity emits a $u$ quark of the
same helicity and subsequently a $c$ quark with the same helicity
as the $u$ quark is re-absorbed forming a $\Lambda_c^+$ with the
same helicity as the proton. On the other hand, the combination
$R_V-R_A$  stands for the situation  where the proton and the
$\Lambda_c^+$ have still the same helicity but the quarks have now
a helicity opposite to that of their parent baryons at one vertex
either at the $p\to\Lambda_c$ or the $\bar{p}\to\bar{\Lambda}_c$
one. The other amplitudes take into account configurations with
either one or two $p\to\Lambda_c^+$ helicity flips.

Strictly speaking, our helicities are light-cone helicities. A
standard c.m.s.\ helicity amplitude is given by the light-cone
helicity amplitude with the same helicities as the c.m.s. one plus
an admixture of other light-cone helicity amplitudes
\be \Phi_{\mu^{\prime} \nu^{\prime}, \mu \nu} =
\frac{1}{1+\beta^2} \left[\mathcal{M}_{\mu^{\prime} \nu^{\prime},
\mu \nu} + \beta \left( 2 \mu^{\prime} \mathcal{M}_{-\mu^{\prime}
\nu^{\prime} , \mu \nu} + 2 \nu^{\prime} \mathcal{M}_{\mu^{\prime}
-\nu^{\prime} , \mu\nu}\right) + 4 \beta^2 \mu^{\prime}
\nu^{\prime} \mathcal{M}_{-\mu^{\prime} -\nu^{\prime}, \mu
\nu}\right]\, . \label{eq:transform} \ee
The strength of the admixture is controlled by the parameter
$\beta$ which reads for our kinematics \ci{diehl01}
\be \beta\= \frac{2M}{(1+\Lambda_M)\sqrt{s}}\,
\frac{\sqrt{tu-t_0u_0}}{\Lambda_M\sqrt{-su}-u-\sqrt{t_0u_0}}\,,
\label{eq:cms-helicity} \ee
where the proton mass is neglected. Near the forward direction
$\beta$ is small and its effect can be ignored, but for larger
scattering angles it is to be taken into account in the
calculation of spin effects. For $M^2\ll s$ \req{eq:cms-helicity}
simplifies to
\be \beta \simeq
\frac{M}{\sqrt{s}}\,\frac{\sqrt{-t}}{\sqrt{s}+\sqrt{-u}}\,. \ee
In this approximation $\beta$ has been used in wide-angle Compton
scattering \ci{HKM}.

%%%%%%%%%%%%%%%%%%%%%%%%%%%%%%%%%%%%%%%%%%%%%%%%%%%%%%%%%%%%%%%%%%%%%%%%
\section{Modelling the GPDs}\label{sec:GPD}
%%%%%%%%%%%%%%%%%%%%%%%%%%%%%%%%%%%%%%%%%%%%%%%%%%%%%%%%%%%%%%%%%%%%%%%

We will model the GPDs by overlaps of LCWFs for the $\Lambda_c^+$ and the
proton. Only the valence Fock states of the baryons are taken into account by
us. For the $\Lambda_c^+$ this is probably a reliable approximation. For the
proton, on the other hand, higher Fock state contributions are likely
important but the required overlap with the $\Lambda_c$ \wf{} projects out
only its valence Fock state.

The overlap formalism for GPDs in terms of LCWFs has been
developed in \ci{DFJK3,brodsky-diehl}. For the process of interest
there is a little complication with it. Due to the unequal mass
kinematics $p_3^\prime$ becomes negative for large scattering
angles in the symmetric frame. The spectator partons, in this
case, move in the direction opposite to their parent hadron. In
order to apply the overlap mechanism in this situation one would
have to boost first into a frame in which $p'_3$ is positive.
Although this can be done in principle, it is fortunately
unnecessary since the $p\bar{p}\to\Lambda_c\bar{\Lambda}_c$ cross
section is strongly peaked in the forward direction for not too
small values of $s$, as will turn out below. Thus, it suffices to
work out the form factors and the cross section only in the
forward hemisphere where $p'_3$ is positive.

In the ERBL region where $\xb<\xi$, meson poles, here $D$ and
$D^*$, may contribute to the GPDs. However, as we discussed in
Sect.\ \ref{sec:peaking}, this region does not contribute to our
process since small $\xb$ are insufficient for the generation of
the $c\bar{c}$ pair. One may also think of Regge exchanges in
general, in addition to the handbag contribution. The most
prominent Regge pole that can contribute to our process is again
the $D^\ast(2007)$. Taking a standard slope for the corresponding
Regge trajectory one finds an intercept of $\alpha(0)=-2.63$ for
it. Since the cross section, say at $t=0$, behaves as
$d\sigma/dt~s^{2(\alpha(0)-1)}$ for a Regge exchange its
contribution is strongly suppressed as compared to the handbag
contribution. Thus, it seems save to neglect pole terms.

In order to calculate the GPDs within the overlap formalism we
need the LCWFs of the proton and the $\Lambda_c$. For the proton a
number of such \wf s  can be found in the literature, e.g.\
\ci{bolz95,bartels07,boffi}. The proton \wf{} advocated for by
Bolz and Kroll (BK) \cite{bolz95} has been determined by fits to
the Dirac form factors of the proton and the neutron at large
momentum transfer, to the valence quark distributions at large $x$
and to the decay width of the process $J/\Psi\to p\bar{p}$. It has
already been applied in an overlap calculation of the proton GPDs
at large $-t$ which form the soft-physics input for an analysis of
wide-angle Compton scattering within the handbag approach
\ci{DFJK1,HKM}. The results of this calculation are found to be in
fair agreement with the data on Compton scattering \ci{bogdan}.
Recently the BK wave function \ci{bolz95} has been supported by
results from lattice QCD\@. The first moments of the proton
distribution amplitude, i.e.\ the LCWF integrated over the quark
transverse momenta, that have been calculated in \ci{lattice}, are
in good agreement with those evaluated from the BK \wf{}. It
should be noted, however, that higher moments tend to weaken the
asymmetry in the distribution amplitude that one would expect from
the first moments only \cite{Braun:2008ur}. The nucleon DA from
the lattice could thus be closer to the asymptotic form of a spin
$1/2$ baryon DA than the nucleon DA resulting from the BK \wf{}.
We think, however, that this is of minor importance for our
purpose of modelling the $u\to c$ transition GPDs and we will use
the BK LCWF in the following.

\begin{figure}[t]
\begin{center}
\includegraphics[width=.40\tw,bb=147 284 474 718,clip=true]{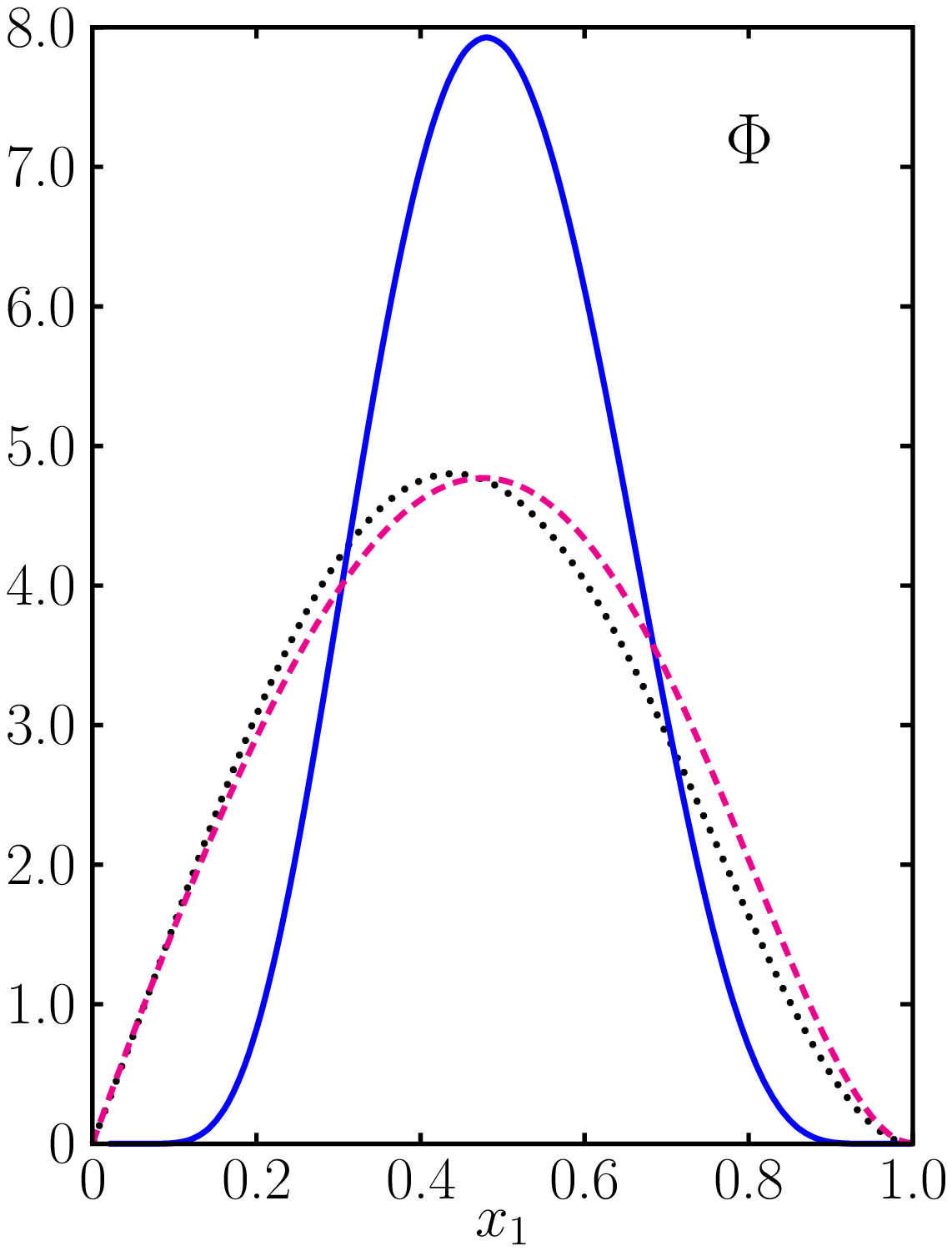}
\end{center}
\caption{The distribution amplitudes \req{BK-DA} (dotted), \req{KK-wf}
  (solid) and the one corresponding to the \wf{} \req{BB-wf} (dashed line)
versus $x_1$ at $x_2=x_3$. (colors online)}
\label{fig:DA}
\end{figure}

The valence Fock state of the proton can be expressed as
\be \mid p,+;uud\rangle \= \int [dx]_3[d^2\vk]_3 \left\{\Psi_{123}
      {\cal M}_{+-+} + \Psi_{213} {\cal M}_{-++}
       - \big(\Psi_{132}+\Psi_{231}\big) {\cal M}_{++-}\right\}\,,
\label{proton-wf}
\ee
with the plane wave exponentials and the color wave function
omitted. The integral measures are defined by
\ba
[dx]_3 &\equiv& dx_1 dx_2 dx_3 \delta(1-\sum_{i=1}^3 x_i)\,,\nn\\
{}[d^2\vk{}]_3  &\equiv& \frac1{(16\pi^3)^2} d^2\vk{}_1 d^2 \vk{}_2 d^2\vk{}_3
                     \delta^{(2)}(\sum_{i=1}^3 \vk{}_i) \,.
\ea
A three-quark state is given by
\be {\cal M}_{\lambda_1\lambda_2\lambda_3}\=
\frac1{\sqrt{x_1x_2x_3}}
               \mid u; x_1,\vk{}_1,\lambda_1\rangle
\mid u; x_2,\vk{}_2,\lambda_2\rangle\,\mid d;
x_3,\vk{}_3,\lambda_3\rangle\,. \label{three-quark-state} \ee
The single particle states satisfy standard normalization.
Eq.~\req{proton-wf} is the most general ansatz for the $L_3=0$
projection of the three-quark proton \wf{}. From the permutation
symmetry between the two $u$-quarks and from the requirement that
the three quarks have to be coupled in an isospin 1/2 state it
follows that there is only one independent scalar \wf{}
\ci{dziembowski} which according to \ci{bolz95} is parameterized
as
\be \Psi^{BK}_{123}(x_i,\vk{}_i)\equiv
\Psi(x_1,x_2,x_3;\vk{}_1,\vk{}_2,\vk{}_3)
         \= N_p (1+3x_1)
         \exp{\left[-a_p^2\sum{\frac{\vk{}_i^2}{x_i}}\right]}\,.
\label{p-wf} \ee
The Gaussian factor is typical for such light-cone \wf s. The
distribution amplitude that corresponds to the \wf{} \req{p-wf} is
\be \Phi_{123}^{BK} \= 60 x_1 x_2 x_3 (1+3x_1)\,. \label{BK-DA} \ee
As usual it is normalized in such a way that $\int[dx]_3\Phi=1$. The
distribution amplitude \req{BK-DA} is displayed in Fig.\
\ref{fig:DA}. It is rather similar to a distribution amplitude
obtained from QCD sum rules \ci{braun-lenz}. The parameters $N_p$
and $a_p$ of the \wf{} \req{p-wf} have been determined in
\ci{bolz95}. They are quoted in Tab.\ \ref{tab:1} together with the
r.m.s.\ value of the quark transverse momentum and the probability
of the proton's valence Fock state evaluated from this \wf{}.

\begin{table*}[t]
\renewcommand{\arraystretch}{1.4}
\begin{center}
\begin{tabular}{|c|| c |c | c | c |c|}
\hline
Ref. & $N$  &  $a$  & $P$  & $\langle \vkk\rangle^{1/2}$ & comment\\[0.2em]
    & $[\gev^{-2}]$ & $[\gev^{-1}]$ &   & $[\mev]$ &\\[0.2em]
\hline
BK \req{p-wf} & 160.5 & 0.75  & 0.17  & 417 &\\[0.2em]
%BM \ci{bartels07} &&&&& \\[0.2em]
 \req{p-qD} & 61.8 & 1.1 &  0.5 & 280   & diquark \\[0.2em]
\hline
KK \req{L-wf}, \req{KK-wf} & 2117  & 0.75  & 0.90 & 460 & $\rho=2.0$ \\[0.2em]
BB \req{L-wf}, \req{BB-wf} & 3477 & 0.75 & 0.90 & 431& $\rho=2.0$ \\[0.2em]
\req{L-qD} & 169.5  & 1.1   & 0.90 & 317  & diquark  \\[0.2em]
\hline
\end{tabular}
\end{center}
\caption{The parameters of the light-cone \wf s used for the calculation
  of the $u\to c$ transition GPDs. For the $\Lambda_c$ the quoted r.m.s.\
  transverse momentum is that of the heavy quark.}
\label{tab:1}
\renewcommand{\arraystretch}{1.0}
\end{table*}

According to the HQET \ci{HQET} the major part of the $\Lambda_c$
state comes from the configuration where the $c$ quark carries the
helicity of its parent baryon while the two light quarks are
coupled in a spin-isospin-zero state. In the  formal limit of
$m_c\to\infty$ there should be no admixture from configurations
where the helicity of the $c$ quark is opposite to that of the
$\Lambda_c$. However, for a $c$-quark mass of $1.27\,\gev$ which
is large enough to justify the application of HQET ideas but too
small in order to suppress corrections to the HQET decisively, a
small admixture from a $c$-quark with helicity opposite to that of
the $\Lambda_c$ cannot be excluded. We therefore write the valence
Fock state of the $\Lambda_c$ in the form
\be \mid\Lambda_c^+,+;cud\rangle\=\int [dx]_3[d^2\vk]_3 \Big\{
({\cal M}^c_{++-}-{\cal M}^c_{+-+}) + \rho (x_2-x_3) {\cal
M}^c_{-++}\Big\}\,\Psi_{\Lambda}(x_i,\vk{}_i)\,,
\label{L-Fock-state} \ee
where ${\cal M}^c_{\lambda_1\lambda_2\lambda_3}$ are the three-quark
states defined in \req{three-quark-state} with the first $u$ quark
being replaced by the $c$ quark. The scalar \wf{}, $\Psi_\Lambda$,
must be symmetric under the interchange of the momentum fractions of
the two light quarks. The factor $x_2-x_3$ in \req{L-Fock-state}
takes care of the isospin of the $\rho$ term \ci{farrar}. It also
guarantees the vanishing of the $\rho$-term in the formal limit
$m_c\to\infty$ as is required by the HQET.

We know from the LEP experiments \ci{ALEPH, OPAL, DELPHI} that the polarization
of the $b$ quark generated in the process $e^+ e^-\to Z\to b\bar{b}$
($P_b=-0.94$) is not fully transferred to the $\Lambda_b$ baryon as is
predicted by the HQET\@. Rather the $\Lambda_b$ polarization only amounts to
$-0.43\pm 0.25$ as an average of the three experimental results. A substantial
fraction of the observed $\Lambda_b$ baryons likely comes from the decay of
the $\Sigma_b$ and $\Sigma^*_b$ which leads to a strong depolarization of the
$\Lambda_b$. Whether this explains the observed $\Lambda_b$ polarization fully
or only partially is unknown. Hence, the possibility that there is an admixture
from configurations where the  $\Lambda_b$ and the $b$ quark have opposite
helicities cannot be ruled out at present. It is plausible to expect that for
the lighter $c$-quark system this configuration is even more important. We
therefore allow for it and take a value of 2.0 for the mixing parameter
$\rho$. The configuration with opposite helicity then contributes about $10\%$
to the probability of the $\Lambda_c$ state \req{L-Fock-state}.

We parameterize the scalar \wf{} in \req{L-Fock-state} as
\be \Psi_{\Lambda} \= N_\Lambda \exp{\big[-f(x_1)\big]}
            \exp{\Big[-a_\Lambda^2\sum \frac{\vk{}_i^2}{x_i}\Big]}\,.
\label{L-wf} \ee
The function $f(x_1)$ has to generate the expected pronounced peak
at $x_1\simeq x_0$ as defined in \req{peak-position}. Current
model \wf s for the $\Lambda_c$ possess this property
\ci{quadder,koerner,ball-braun,guo01,BSW,guo91}. For $f(x_1)$ we
adopt a slightly modified version of the one given in
\cite{koerner},
\be f_{KK}(x_1) \= a_\Lambda^2 M^2
\frac{(x_1-x_0)^2}{x_1(1-x_1)}\,. \label{KK-wf} \ee
The LCWF \req{L-wf} is a suitable adaptation of
the one proposed in \ci{BSW} which has been obtained by transforming
a harmonic oscillator \wf{} to the light cone \ci{guo91}.

As an alternative we will also make use of a result from QCD sum
rules obtained for the $\Lambda_b$ baryon \ci{ball-braun} which we
adapt to our case of a charm baryon, namely
\be f_{BB}(x_1) \= a_\Lambda M  (1-x_1)\,. \label{BB-wf} \ee
Up to corrections of order $(M-m_c)/M$ the two functions have the
same $x_1$ dependence for large $m_c$. The parameters of the two
choices for the $\Lambda_c$ \wf{} are quoted in Tab.\ \ref{tab:1}.
We assume $a_p=a_\Lambda$ and a probability of $0.9$. The
normalization $N_\Lambda$ is fixed from these parameters. A large
value of the $\Lambda_c$ valence Fock state probability is to be
expected~\ci{BSW}.

The distribution amplitude corresponding to \req{L-wf} reads
\be \Phi^a_\Lambda \propto x_1 x_2 x_3 \exp{\big[-f(x_1)\big]}\,.
\label{L-DA} \ee
The two $\Lambda_c$ distributions amplitudes are compared to each
other and to the proton distribution amplitude in Fig.\
\ref{fig:DA}. The variant \req{BB-wf} is broader than \req{KK-wf}
and is very similar to the proton distribution amplitude. Since
the $c$ quark is relatively light the peak positions of the
$\Lambda_c$ and the proton distribution amplitudes do not differ
much. For the case of the $\Lambda_b$, for instance, the peak
positions would differ markedly.

The GPDs are obtained from the following overlap integrals \ci{DFJK3}
\ba H^{cu}_T(\xb,\xi,t)&=&- \frac1{1-\xi^2}\int
[d\xb]_3[d^{\,2}\vk]_3\delta(\xb-\xb_1)
          \Psi_\Lambda(\hat{x}_i,\hat{k}_{\perp i}) \nn\\
     &\times&       \Big[\Psi_{123}(\tilde{x}_i,\tilde{k}_{\perp i})
                   +\Psi_{132}(\tilde{x}_i,\tilde{k}_{\perp i})
                   +\Psi_{231}(\tilde{x}_i,\tilde{k}_{\perp i})\Big] \,, \nn\\
\Delta H^{cu}(\xb,\xi,t)&=& -\frac{\rho}{1-\xi^2}\int [d\xb]_3[d^{\,2}\vk]_3\delta(\xb-\xb_1)
               \Psi_\Lambda(\hat{x}_i,\hat{k}_{\perp i}) \nn\\
          &\times&  (\hat{x}_2-\hat{x}_3)
                   \Psi_{213}(\tilde{x}_i,\tilde{k}_{\perp i})\,,
\label{eq:Hqqq}
\ea
and the GPDs $H$ and $\widetilde{H}$ are given by the combinations
\be H^{cu}\=H^{cu}_T - \Delta H^{cu}\,, \qquad
\widetilde{H}^{cu}\=H^{cu}_T + \Delta H^{cu}\,.
\label{eq:H-structure} \ee
All other GPDs are zero. They require parton orbital angular
momentum, a property that our simple \wf s for the $\Lambda_c$ do
not possess. At least for large $m_c$ its structure follows from
the HQET. The proton \wf s might have a more complicated structure
involving, for instance, components with non-zero orbital angular
momentum. This would still lead to the same result for the GPDs
since the overlap with the $\Lambda_c$ LCWF does project out only
$L=0$ components of the proton \wf.

\begin{figure}[t]
\begin{center}
\includegraphics[width=0.415\tw,bb= 126 296 461 727,clip=true]{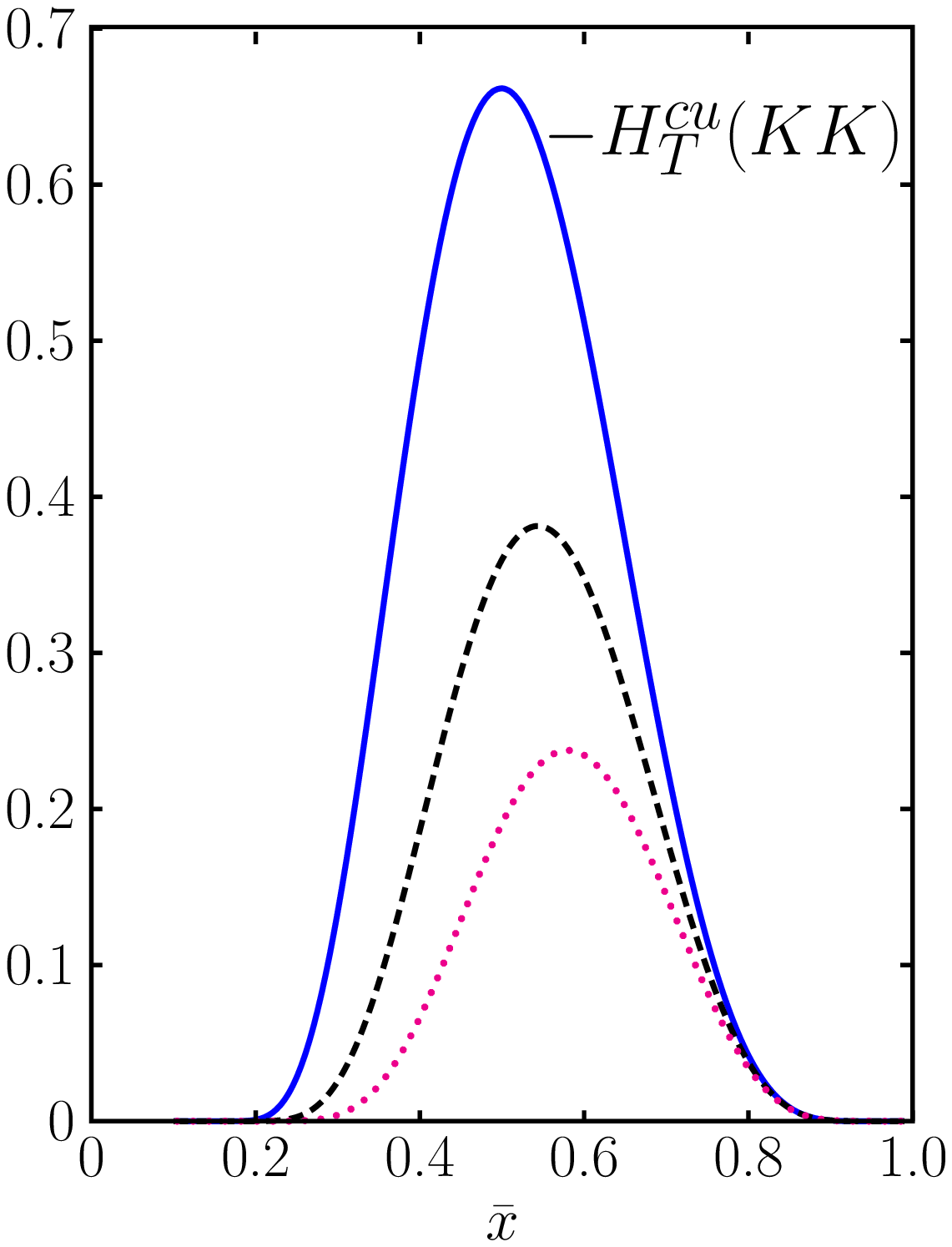}
\includegraphics[width=0.415\tw,bb= 126 296 461 727,clip=true]{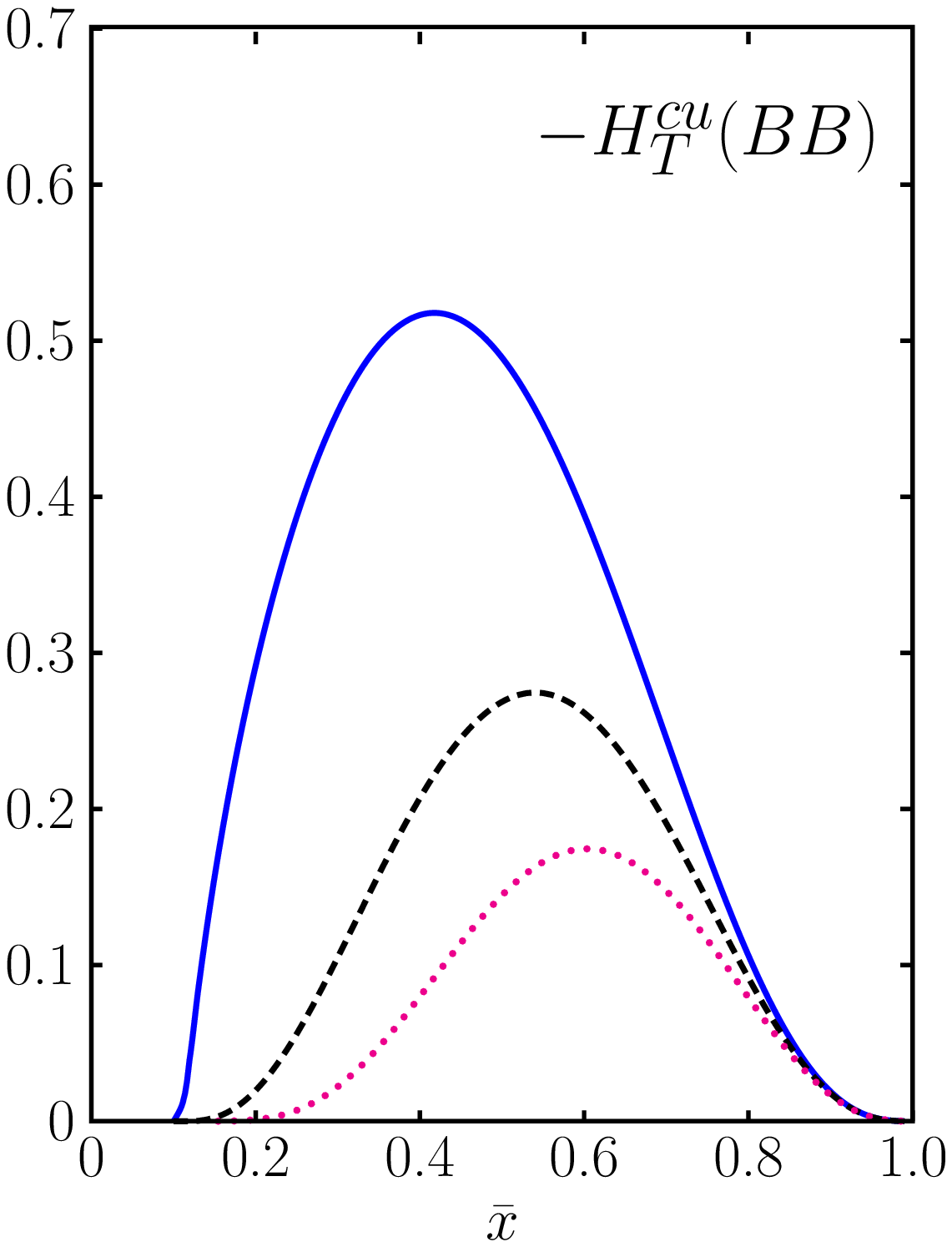}
\end{center}
\caption{The $u\to c$ transition GPD $H_T$ obtained from the
overlap of the LCWFs
  \req{p-wf} and \req{L-wf} (for the variants \req{KK-wf} (left) and
  \req{BB-wf} (right)) versus $\xb$ at $s=30\,\gev^2$ and
$\vdd=0, 2.0, 4.0\,\gev^2$, corresponding to
$t^\prime=0,-2.17, -4.41\,\gev^2$ ($\xi= 0.112,  0.123, 0.139$) shown as
solid, dashed and dotted lines, respectively (colors online).}
\label{fig:GPD-qqq}
\end{figure}

The arguments of the light-cone \wf {s} are related to the average parton
momenta, over which it is to be integrated, by \ci{DFJK3}
\ba \tilde{x}_1 &=& \frac{\xb_1+\xi}{1+\xi}\,, \qquad
   \tilde{k}_{\perp 1}\= \bar{k}_{\perp 1}-\frac{1-\xb_1}{1+\xi}\frac{\vd}{2}\,,\nn\\
\tilde{x}_j &=& \frac{\xb_j}{1+\xi}\,, \qquad \;
  \tilde{k}_{\perp j}\= \bar{k}_{\perp j}+\frac{\xb_j}{1+\xi}\frac{\vd}{2}\,,
 \quad {\rm for}\;\; j=2,3, \nn\\
\hat{x}_1 &=& \frac{\xb_1-\xi}{1-\xi}\,, \qquad
   \hat{k}_{\perp 1}\= \bar{k}_{\perp 1}+\frac{1-\xb_1}{1-\xi}\frac{\vd}{2}\,,\nn\\
\hat{x}_j &=& \frac{\xb_j}{1-\xi}\,, \qquad\;
  \hat{k}_{\perp j}\= \bar{k}_{\perp j}-\frac{\xb_j}{1-\xi}\frac{\vd}{2}\,,
 \quad {\rm for}\;\; j=2,3.
\ea
Working out these integrals for the \wf s \req{p-wf} and \req{L-wf} we obtain
the GPD $H^{cu}_T$
\ba H^{cu}_T &=& -\frac{3}{4} \frac{N_\Lambda
N_p}{(4\pi)^4}\frac1{1-\xi^2}
        \frac{\xb^2-\xi^2}{a^2_p(1+\xi)+a^2_\Lambda(1-\xi)}\nn\\
    &\times& \frac{(1-\xb)^3}{1+\xi}\frac{1+2\xi+\xb}
     {a_p^2(1+\xi)^2(\xb-\xi)+a^2_\Lambda(1-\xi)^2(\xb+\xi)}\nn\\
  &\times&
\exp{\Big[-f\Big(\frac{\xb-\xi}{1-\xi}\Big)\Big]}\,
           \exp{\Big[\frac{-(1-\xb)\,a^2_\Lambda a^2_p\,\vdd}
        {a_p^2(1+\xi)^2(\xb-\xi)+a^2_\Lambda(1-\xi)^2(\xb+\xi)}\Big]}\,,
\label{eq:Hqqq-expl} \ea
and for the $\rho$-dependent part
\be \Delta H^{cu} \=
\frac{\rho}{15}\frac1{1-\xi}\,\frac{(1-\xb)^2}{1+2\xi+\xb}\,H^{cu}_T\,.
\label{eq:Hqqq-expl-rho} \ee
The GPD $H_T$ is shown in Fig.\ \ref{fig:GPD-qqq}. One observes
that the GPD exhibits the anticipated pronounced peak at
$\xb\simeq x_0$. With increasing $\vdd$ (or $-t^\prime$, see
Eq.~(\ref{def:tprime})) the position of the peak is slightly
shifted towards larger values of $\xb$. This property of the $u\to
c $ transition GPDs is similar to that of the proton valence-quark
GPDs which have been determined from an analysis of the nucleon
form factors through the familiar GPD sum rules in Ref.\
\ci{DFJK4}. The only difference is that the $u\to c$ transition
GPDs possess this property already at $t^\prime=0$ as a
consequence of the intrinsic scale $4m_c^2$. The proton GPDs
exhibit a peak only at large values of $-t$ and, therefore, the
handbag approach can be applied to (real) Compton scattering only
at wide angles. The two variants of the mass exponential entail
striking differences in shape of the resulting GPDs displayed in
Fig.\ \ref{fig:GPD-qqq}. The GPD evaluated from \req{BB-wf} is
much broader than the other one. The maxima of the GPD evaluated
from \req{BB-wf} are sited at lower values of $\xb$ for small
$\vdd$ but at slightly larger ones for large $\vdd$ than is
obtained from \req{KK-wf}. The differences between the two GPDs
have a bearing on the form factors and predicted cross sections as
we are going to discuss in the following.

\begin{figure}[t]
\begin{center}
\includegraphics[width=0.39\tw,bb= 126 232 450 695,clip=true]{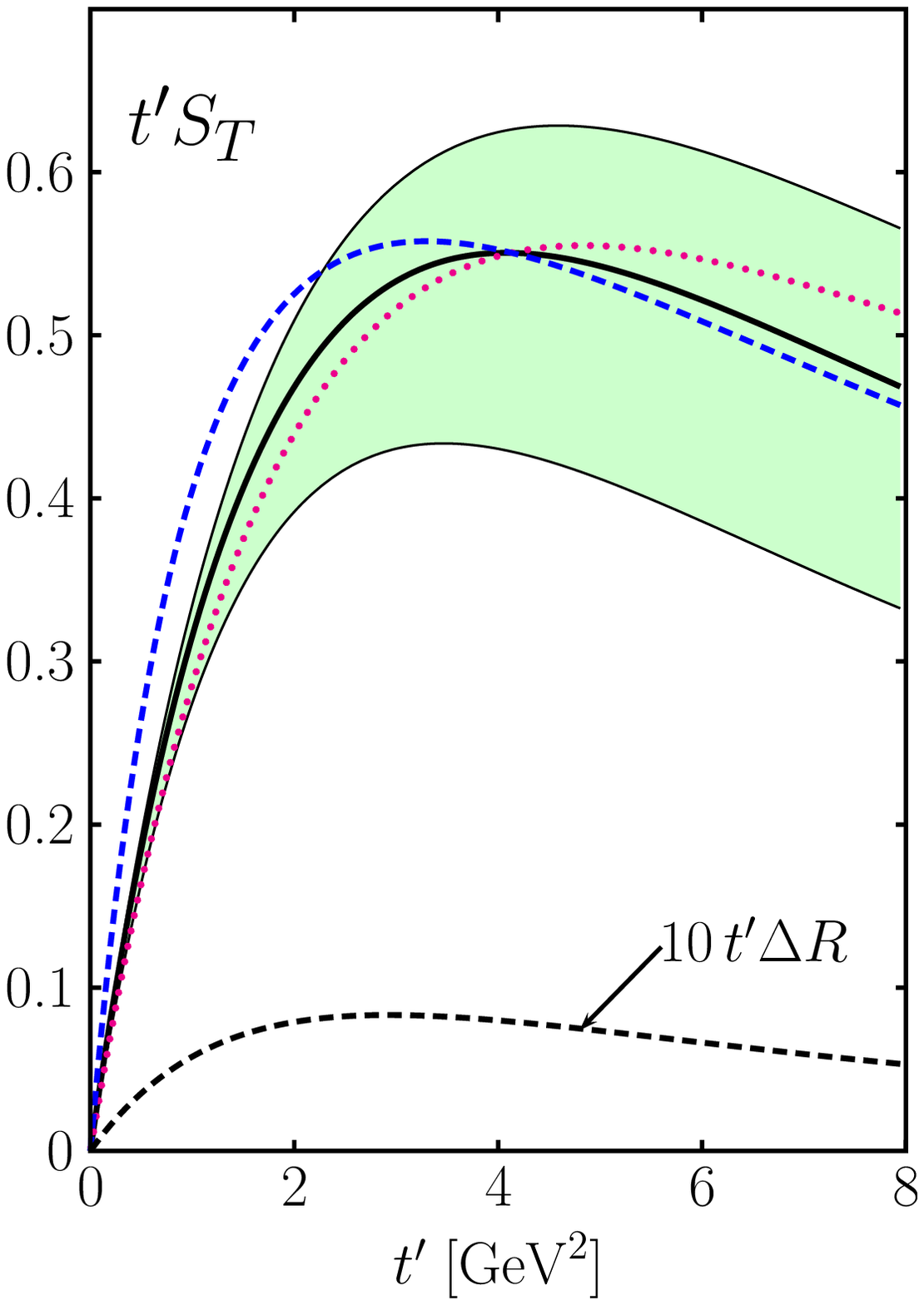}
\end{center}
\caption{The transition form factor $S_T$ scaled by $t^\prime$ versus $t^\prime$
at $s=30\,{\rm GeV}^2$ evaluated from \req{eq:Hqqq-expl} using \req{KK-wf}.
The band indicates the parametric uncertainties of the form factor. The dashed
line represents the the result obtained with \req{BB-wf} instead of
\req{KK-wf} and the dotted line the approximation \req{eq:FFappr}. The lower
dashed line is 10 times the corresponding scaled form factor $\Delta R$ (colors online).}
\label{fig:FF-qqq}
\end{figure}

With the GPDs at hand the transition form factors can be evaluated
according to \req{eq:annFF}. They exhibit a structure analogously to
\req{eq:H-structure}
\be R_V(\xi,t)\=S_T(\xi,t)-\Delta R(\xi,t)\,, \quad
R_A(\xi,t)\=S_T(\xi,t)+\Delta R(\xi,t)\,. \label{eq:R-structure}
\ee
In Fig.\ \ref{fig:FF-qqq} the form factors $S_T$ and $\Delta R$,
scaled by $t^\prime$ for the ease of graphical representation, are
shown at $s=30\,\gev^2$. The differences between the models
\req{KK-wf} and \req{BB-wf} are particularly substantial at small
$\vdd$ or $-t^\prime$. Also displayed in Fig.\ \ref{fig:FF-qqq} is
the approximation \req{eq:FFappr}. As can be observed it is very
close to the full form factor. This approximate coincidence nicely
demonstrates the internal consistency of the peaking
approximation. As the figure also reveals, $\Delta R$ is tiny
despite the fact that the $\rho$ term in \req{L-Fock-state}
contributes as much as $10\%$ to the probability of the
$\Lambda_c^+$ state. The strong suppression of $\Delta R$ is a
consequence of the tiny overlap of the $\rho$ term in the
$\Lambda_c$ state with the proton \wf{} which manifests itself in
the factor $(1-\xb)^2$ in \req{eq:Hqqq-expl-rho}. Even by a
doubling of $\rho$, which leads to the implausibly large
contribution of $30\%$ to the probability of the $\Lambda_c$
valence Fock state, we obtain $\mid \Delta R\mid \ll \mid
S_T\mid$. Thus, $\Delta R$ is so small that it has no direct
bearing on the observables for realistic values of $\rho$.

Let us now turn to the issue of the error assessment. It is true
that the independent parameters of the $\Lambda_c$ LCWF are the
normalization and the transverse size parameter, but we do not have
any idea about their uncertainties in contrast to the probability
and the r.m.s.\ transverse momentum. We, therefore, assume errors
for the latter quantities and vary $N_\Lambda$ and $a_\Lambda$ as
well as $m_c$ (within the PDG limits \req{c-error}) in such a way
that the errors of $P_\Lambda$ and $\langle \vkk\rangle^{1/2}$ are
covered. For the latter quantity we assume an error of $\pm 10\%$
which is in fair agreement with an estimate of it for heavy baryons
made in \ci{guo01}. For the probability of the $\Lambda_c$ valence
Fock state we allow for the maximum value of 1 and take as its
lowest value 0.7. The described estimate of the parametric error is
shown for the form factor $S_T$ in Fig.\ \ref{fig:FF-qqq}. We do not
take into account the uncertainties of the proton \wf{}. They are
small compared to that of the $\Lambda_c$ \wf{} since the proton
\wf{} is determined from detailed fits to nucleon form factors and
parton distributions \ci{bolz95}. The uncertainty of the form factor
evaluated from the mass exponential \req{BB-wf} is of about the same
size as for the variant \req{KK-wf}.

It is popular to model the baryons as a quark-diquark state. For
the case of the $\Lambda_c$ this appears rather natural with
regard to the HQET and indeed many examples of quark-diquark \wf s
for the $\Lambda_c$ can be found in the literature
\ci{quadder,koerner,guo01,chinese,ebert,hernandez}. In all these
cases only a spin-isospin scalar diquark, $S_{[ud]}$, is used, an
assumption that is in accordance with the large-$m_c$ limit of the
HQET\@. In view of our experience with the three-quark models we
omit a possible admixture of other diquarks for simplicity. The
proton may have a more complicated spectator structure with scalar
and axial-vector diquarks and perhaps excitations of higher states
involving angular momentum \ci{schweiger}. But as we mentioned in
the discussion of the 3-quark models, the overlap with the
$\Lambda_c^+$ LCWF only projects out the corresponding component
with the scalar diquark from the proton LCWF\@. In order to render
possible a comparison with the results presented in \ci{quadder}
we also employ the quark-diquark model here. For the ease of
comparison we use the same parameters as in \ci{quadder} although
the LCWFs are slightly modified in order to take care of recent
developments.

According to \cite{quadder} we write the \wf{} of a proton in an
$u_{\pm}S_{[ud]}$ Fock state as
\be \Psi_{p(qD)} \= N_{p(qD)} (1-x)
\exp{\big[-\frac{a_{p(qD)}^2\vk^2}{x(1-x)}\big]}\,. \label{p-qD}
\ee
The parameters of this \wf{} are quoted in Tab.\ \ref{tab:1}. The
probability of 0.5 for a proton may appear rather large. However,
one may argue that a diquark is a bound state of two quarks and
embodies therefore also gluons and sea quarks. Hence, the
quark-diquark \wf{} effectively takes into account higher Fock
states and has therefore a larger probability than one would
expect for a 3-quark valence-Fock-state \wf{}. In view of this
fact a larger transverse size of the quark-diquark state appears
plausible, too.

For the LCWF of a $\Lambda_c^+$ in a $c_{\pm}S_{[ud]}$
state we use
\be \Psi_{\Lambda (qD)}\=N_{\Lambda (qD)} (1-x)
            \exp{\Big[-a_{\Lambda (qD)}^2M^2\frac{(x-x_0)^2}{x(1-x)}\Big]}
            \exp{\Big[-\frac{a_{\Lambda (qD)}^2 \vk^2}{x(1-x)}\Big]}\,.
\label{L-qD}
\ee
The parameters are again listed in Tab.\ \ref{tab:1}.

The overlap of the two wave functions \req{p-qD} and \req{L-qD} provides the
following result for $H=\widetilde{H}=H_T$
\ba H^{cu}_T&=& \frac{N_{\Lambda (qD)} N_{p (qD)}}{16\pi^2}
\frac1{(1-\xi^2)^{3/2}}
    \frac{(1-\xb)^3(\xb^2-\xi^2)}
     {a_{\Lambda (qD)}^2(1-\xi)^2(\xb+\xi)+a_{p(qD)}^2(1+\xi)^2(\xb-\xi)}\nn\\
 &\times& \exp{\Big[-a_{\Lambda (qD)}^2 M^2\frac{(\xb-\xi-x_0(1-\xi))^2}{(\xb-\xi)(1-\xb)}\Big]}\nn\\
 &\times& \exp{\Big[\frac{-(1-\xb)a_{\Lambda (qD)}^2 a_{p(qD)}^2\,\vdd}
      {a_{\Lambda (qD)}^2(1-\xi)^2(\xb+\xi)+a_{p(qD)}^2(1+\xi)^2(\xb-\xi)}\Big]}\,.
\ea
The other GPDs are zero. The quark-diquark GPD has a similar shape
as the 3-quark GPDs. Only its size is larger by about a factor of 1.6. The
form factors $R_V=R_A=S_T$ in the quark-diquark model also behave similar to those
in the 3-quark model.

%%%%%%%%%%%%%%%%%%%%%%%%%%%%%%%%%%%%%%%%%%%%%%%%%%%%%%%%%%%%%%%%%%
\section{Observables}
\label{sec:observables}
%%%%%%%%%%%%%%%%%%%%%%%%%%%%%%%%%%%%%%%%%%%%%%%%%%%%%%%%%%%%%%%%%%%
\subsection{Cross sections}
%%%%%%%%%%%%%%%%%%%%%%%%%%%%%%%%%%%%%%%%%%%%%%%%%%%%%%%%%%%%%%%%%%
The differential cross section for the process
$p\bar{p}\to\Lambda_c^+\bar{\Lambda}_c^-$ reads
\be \frac{d\sigma}{d\Omega}\=\frac1{4\pi} s\Lambda_m\Lambda_M
\frac{d\sigma}{dt}
\=\frac{1}{64\pi^2}\,\frac1{s}\frac{\Lambda_M}{\Lambda_m}
\sigma_0\,, \ee where \be \sigma_0 \= \frac14\sum \left |{\cal
M}_{\mu'\nu',\mu\nu}\right |^2\,. \ee
The cross section can readily be calculated from the amplitudes
\req{eq:lc-amplitudes} using the subprocess amplitudes given in
\req{subprocess-amp} and appropriate form factors. Particularly
simple are those GPD models in which the $\Lambda_c$ only consists
of the configuration where the two light quarks are coupled in a
spin-isospin-zero state (bound in a diquark or not) as is
predicted by the HQET\@. They do not involve quark orbital angular
momentum and the parameter $\rho$ is zero. As we discussed in the
preceding section a non-zero value of $\rho$ is possible but
cannot be large as estimated from measurements of the $\Lambda_b$
polarization at LEP \ci{ALEPH,OPAL,DELPHI}. In the cases we
examine, the $\rho$ term does only lead to tiny differences in the
form factors. Practically, there is only one independent form
factor
\be S_T(\xi,t)\simeq R_V(\xi,t)\simeq R_A(\xi,t)\,.
\label{eq:FF-relation} \ee
All other form factors are zero. In a situation where
\req{eq:FF-relation} holds, the differential cross section
strongly simplifies and becomes proportional to the subprocess one
\be \frac{d\sigma}{d\Omega}(p\bar{p} \to \Lambda_c^+
\bar{\Lambda}_c^-)
          \= \left(\frac{C_F}{N_C}\right)^2 (1-\xi^2)^2
\Big(\frac{m_c}{M}\Big)^2 \frac{S_T^{\,4}(\xi,t)}{\Lambda_m}\;
\frac{d\hat{\sigma}}{d\hat{\Omega}}(\ubu\to \cbc)\,.
%\sigma &=& \left(\frac{C_F}{N_C}\right)^2 (1-\xi^2)^2 \Big(\frac{m_c}{M}\Big)^2
%               \frac{R_V^{\,4(t)}}{\Lambda_m}\; \hat{\sigma} (\ubu\to \cbc)\,,
\ee

\begin{figure}[ht]
\begin{center}
 \includegraphics[width=.40\textwidth,bb=110 287 473 731,%
clip=true]{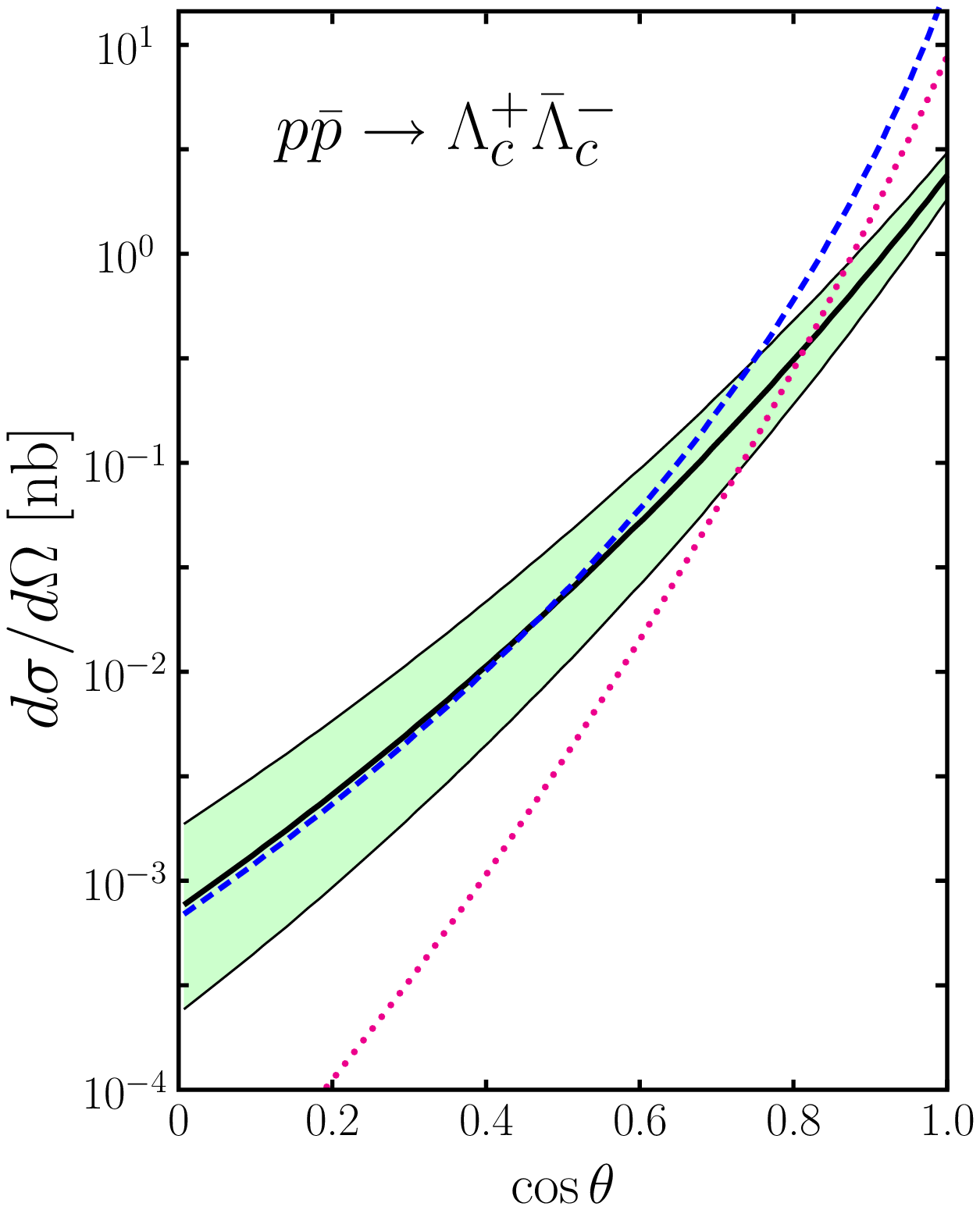}
 \includegraphics[width=.40\textwidth,bb=111 287  473 731,%
clip=true]{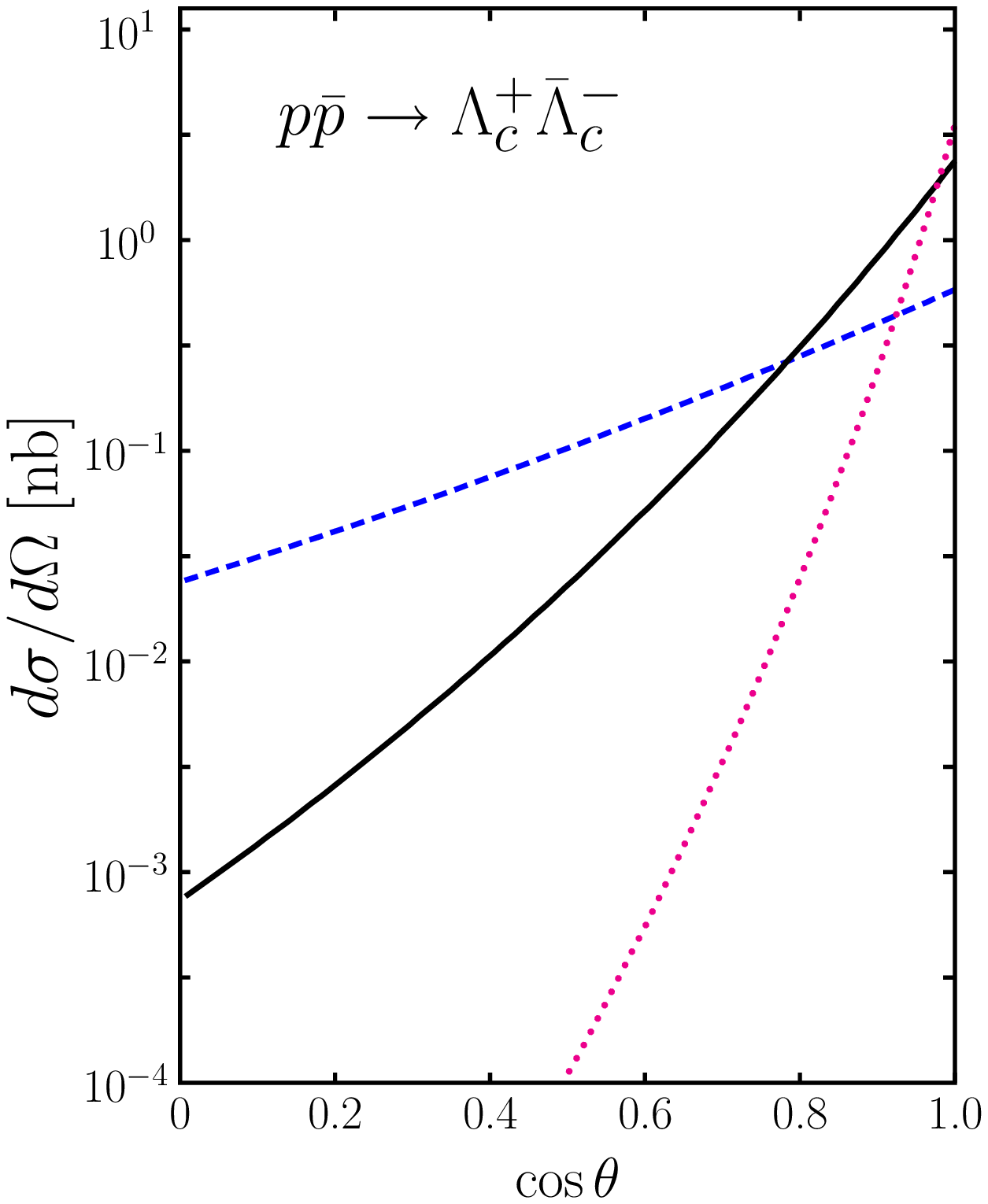}
\end{center}
\caption{Predictions for the differential cross section versus $\cos{\theta}$.
    Left: $s=30\,\gev^2$. Solid line represents the result evaluated
    from \req{eq:Hqqq-expl} with the mass exponential \req{KK-wf}
    (with error band). The dashed line is obtained with \req{BB-wf} and the
    dotted one is the results from the quark-diquark model. Right: The
    differential cross section at $s=23\,, 30$ and  $50\,\gev^2$ shown as
    dashed, solid and dotted line, respectively. The cross sections are
    evaluated from \req{eq:Hqqq-expl} with the mass exponential \req{KK-wf}
    (colors online).}
\label{fig:dsdo}
\end{figure}

Results for the differential cross sections are shown in Fig.\
\ref{fig:dsdo}. The differential cross section is sharply forward
peaked. This effect becomes more pronounced for higher energies.
It arises from the dominance of the amplitude ${\cal M}_{+-,+-}$
($\propto \cos^2{\theta/2}$) near the forward direction and is
natural given that we consider an annihilation reaction at high
energies. The effect is pronounced by the behavior of the form
factors which obey \req{eq:FF-relation} and decrease monotonically
with increasing $-t^\prime$. The cross section increases with $s$
for small scattering angles near the forward direction. This is a
transient effect which disappears at large $s$ and
$d\sigma/d\Omega$ ultimately show the usual $1/s$ decrease since
the form factors become independent on $\xi$ in that region as we
mentioned in Sect.\ \ref{sec:peaking}. The various models we
employ behave similar in that respect. The quark-diquark model
predicts a particularly steeply falling cross section. This
feature is only to be expected given the small values of the
r.m.s.\ transverse momenta in that model or the associated large
transverse sizes (see Tab.~ \ref{tab:1}). It can also be seen in
Fig.~\ref{fig:dsdo} that the use of the mass exponential
\req{BB-wf} instead of \req{KK-wf} provides markedly larger
differential cross sections for small-angle scattering.

\begin{figure}[t]
\begin{center}
\includegraphics[width=.50\textwidth,bb=126 407 519 733,%
clip=true]{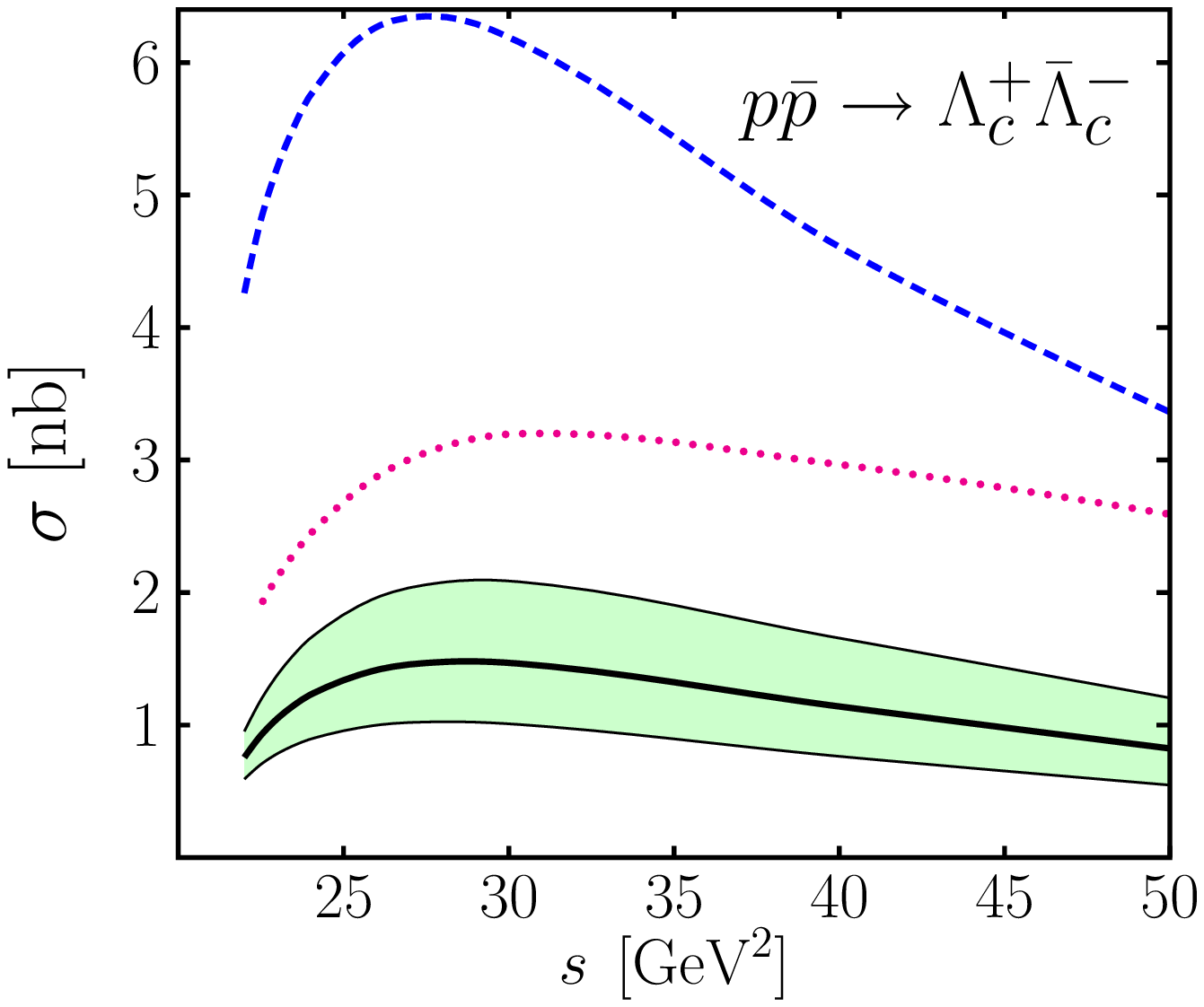}
\end{center}
\caption{The integrated cross section ($\cos\theta \ge 0$) versus
$s$. For notation we refer to the left hand side of Fig.\
\ref{fig:dsdo}.} \label{fig:qqq-cross-section}
\end{figure}

The integrated cross section is displayed in Fig.\
\ref{fig:qqq-cross-section}. Its magnitude is of order nb. As the
behavior of the differential cross section indicates (see Fig.\
\ref{fig:dsdo}) the largest cross section is obtained with the mass
exponential \req{BB-wf}. For comparison we note that in \ci{quadder}
the $p\bar{p}\to\Lambda_c^+\bar{\Lambda}_c^-$  cross section was
estimated to amount to about 3(10) nb at $s=30 (50)\, \gev^2$. This
is in reasonable agreement with the predictions from the
quark-diquark model as well as the other ones presented here.

In Sect.\ \ref{sec:GPD} we have already discussed the model uncertainties of
the form factors. These uncertainties enter directly into the errors of the
cross section (note that it is related to the fourth power of the form
factor). The errors are shown as bands in Figs.\ \ref{fig:dsdo} and
\ref{fig:qqq-cross-section}. The width of the bands correspond to a
probability of the relevant $\rho$-independent term in the $\Lambda_c$ state
\req{L-Fock-state} varying between 0.9 and 0.6 and to an error of $\pm 10\%$
for the r.m.s.\ charm-quark transverse momentum. All the models we exploit have
similar parametric errors. Larger than these parametric errors is
the difference between the predictions obtained from the different mass
exponentials which is rather representative for the uncertainties of our
predictions.

%%%%%%%%%%%%%%%%%%%%%%%%%%%%%%%%%%%%%%%%%%%%%%%%%%%%%%%%%%%%%%%%%%%%%%%%%%%%%
\subsection{Spin dependence}
%%%%%%%%%%%%%%%%%%%%%%%%%%%%%%%%%%%%%%%%%%%%%%%%%%%%%%%%%%%%%%%%%%%%%%%%%%%%
For spin-dependent observables we have to take care of the fact that
experiments provide typically c.m.s.\ observables which are
expressed in terms of c.m.s.\ helicity amplitudes ($\Phi$). We
therefore have to transform our light-cone amplitudes
\req{eq:combi}, \req{eq:lc-amplitudes} to the c.m.s.\ ones
\req{eq:transform} with the help of the parameter $\beta$ given in
\req{eq:cms-helicity}. We note that for observables that only refer
to the helicities of the initial state baryons this transform has no
effect up to corrections of order $m/\sqrt{s}$. We also remark that
the unpolarized cross section can be worked out in either basis, it
is independent on the transform \req{eq:transform}.

While all single-spin asymmetries are zero to lowest order of
perturbative QCD, there are many non-zero spin correlation
parameter in our approach. We employ the conventions for the
notation of such observables advocated for in \ci{soffer}. The
directions of the spin of the particles are denoted by $L \dots$
helicity, $N \dots$ normal to scattering plane, $S \dots$
`sideways' spin direction within scattering plane but orthogonal
to direction of momentum,
\be {\bf L}_i \= {\bf p}_i/{p_i}\,, \qquad {\bf N} \= {\bf L}_p
\times {\bf L}^\prime_\Lambda\,, \qquad {\bf S}_i \= {\bf N}\times
{\bf L}_i\,. \ee
The corresponding spin states for $N$ and $S$ type polarizations
read
\be \mid \uparrow (\downarrow)\rangle \=
\frac1{\sqrt{2}}\,\Big[\mid+\rangle +(-) \imath
\mid-\rangle\Big]\,,\qquad \mid \to (\gets)\rangle \=
\frac1{\sqrt{2}}\,\Big[\mid+\rangle +(-)  \mid-\rangle\Big]\,, \ee
respectively. The spin-spin correlation parameters are denoted by ($i,j=L,N,S$):\\
%first label refers to first particle)
\hspace*{0.1\tw} initial state spin correlations  \hspace*{0.24\tw}$\quad - \quad A_{ij}$,\\
\hspace*{0.1\tw} final state spin correlations \hspace*{0.25\tw}  $\quad - \quad C_{ij}$,\\
\hspace*{0.1\tw} polarization transfer from $p(\ov{p})$ to $\Lambda_c^+
(\ov{\Lambda}{}_c^-)$ \hspace*{0.085\tw} $\quad - \quad D_{ij} (\ov{D}_{ij})$, \\
\hspace*{0.1\tw} polarization correlation between $p(\ov{p})$ and $\ov{\Lambda}{}_c^-
(\Lambda_c^+)$  $\quad - \quad K_{ij} (\ov{K}_{ij})$.\\

A spin correlation parameter is defined by
\be {\cal O}_{ij} \= \frac{\sigma(ij)+\sigma(-i-j)-\sigma(i-j)
-\sigma(-ij)}
                 {\sigma(ij)+\sigma(-i-j)+\sigma(i-j) +\sigma(-ij)}\,,
\ee
where $\sigma(ij)$ stands for the differential cross section for
the scattering with two particles in given spin states. For all
GPD models that approximately possess the property
\req{eq:FF-relation}, the  spin correlations simply
read~\footnote{ Of course, the subprocess amplitudes have to be
transformed to the c.m.s.\ basis analogously to \req{eq:transform}
and \req{eq:cms-helicity}.}
\be {\cal O}(p\bar{p} \to \Lambda_c^+ \bar{\Lambda}_c^+) \= {\cal
O} (\ubu\to \cbc)\,. \ee
The form factors and, hence, their uncertainties, cancel out in
the spin correlations. Thus, all our GPD models lead to the same
set of predictions for these observables. In this sense our
predictions for spin correlations are model independent.

\begin{figure}[t]
\begin{center}
\includegraphics[width=.40\textwidth,bb=145 328 473 709,%
clip=true]{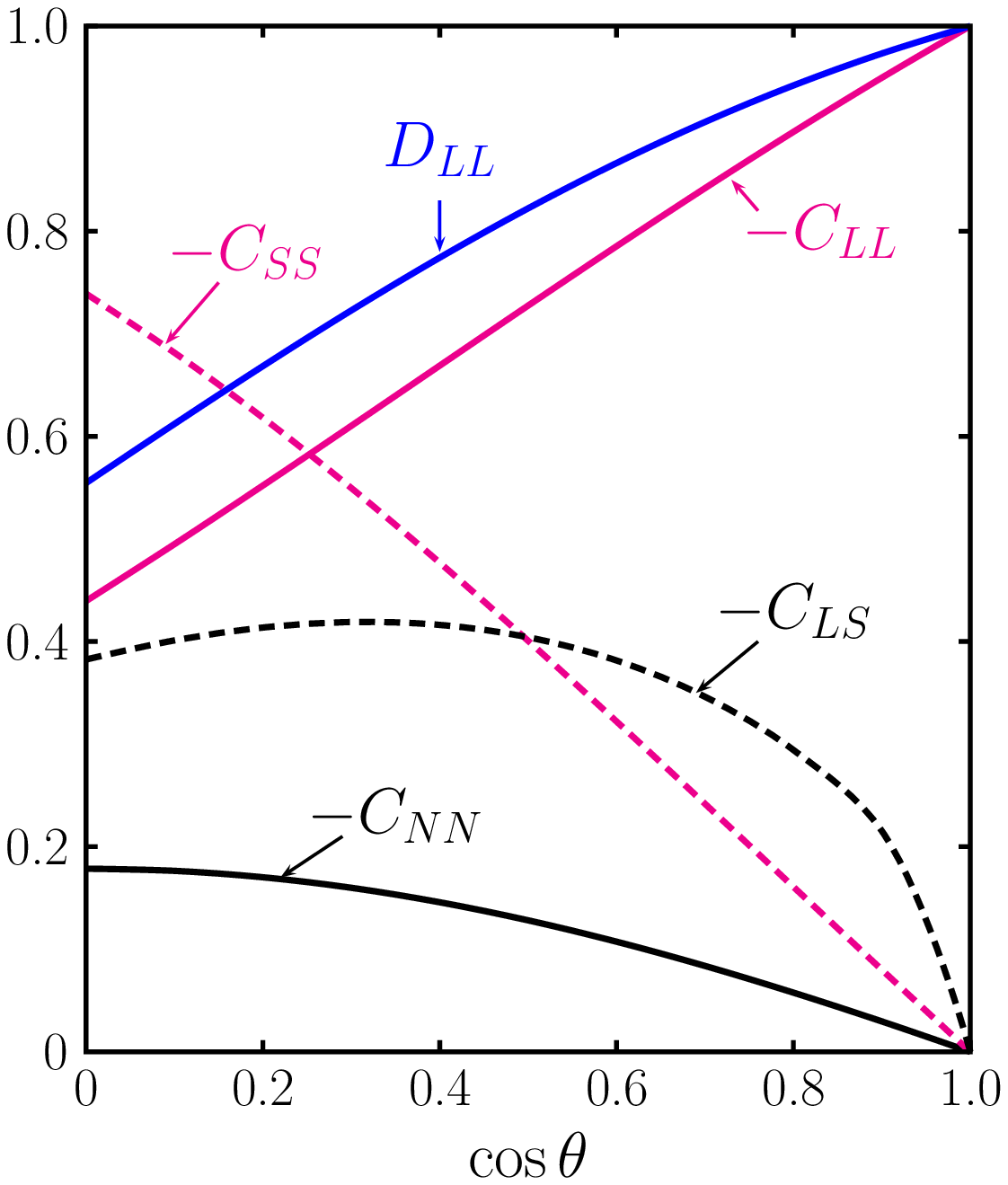}
\end{center}
\caption{Final state helicity correlations $C_{ij}$ and the $p\to \Lambda_c$
polarization transfer $D_{LL}$ versus $\cos{\theta}$ at $s=30\,\gev^2$.
(colors online)}
\label{fig:spin}
\end{figure}

Below we list a set of examples of spin correlations obtained in
the handbag approach from form factors which satisfy the relation
\req{eq:FF-relation}. If the parameter $\beta$ is only taken into
account up to linear order, which is justified for not too large
scattering angles, the corresponding analytical expressions become
rather simple:
\ba
A_{LL} &=& -1\,,   \nn\\
C_{LL} &=& - \frac{1+\cos^2{\theta} -4M^2/s\sin^2{\theta}
  +8\beta M/\sqrt{s}\sin{2\theta}}
{1+\cos^2{\theta}  +4M^2/s\sin^2{\theta}}\,,\nn\\
D_{LL}&=& -K_{LL} \=2 \frac{\cos{\theta} + 4\beta M/\sqrt{s}\sin{\theta}}
          {1+ \cos^2{\theta} + 4M^2/s\sin^2{\theta}}\,, \nn\\
D_{NN} &=& D_{SS}\=0\,, \nn\\
%A_{NN} &=& -\sin{\theta}\,\frac{\sin{\theta}(1-4M^2/s)-16\beta M/\sqrt{s}\cos{\theta}}
%              {1+ \cos^2{\theta} + 4M^2/s\sin^2{\theta}}\,, \nn\\
A_{NN} &=& C_{NN} \= -\sin^2{\theta}\,\frac{1-4M^2/s}
           {1+ \cos^2{\theta} + 4M^2/s\sin^2{\theta}}\,, \nn\\
C_{SS} &=& -\sin{\theta}\,\frac{\sin{\theta}(1+4M^2/s) - 16\beta
              M/\sqrt{s}\cos{\theta}}
            {1+ \cos^2{\theta} + 4M^2/s\sin^2{\theta}}\,,\nn\\
C_{LS} &=& -4\frac{M/\sqrt{s}\sin{\theta}\cos{\theta}-\beta\cos^2{\theta}
            +4\beta M^2/s\sin^2{\theta}} {1+ \cos^2{\theta} + 4M^2/s\sin^2{\theta}}\,,\nn\\
C_{LN} &=& C_{NS} \=0\,. \ea
As for single spin asymmetries (cf.\ the remark in Sect.\
\ref{sec:subprocess}) the observables $D_{NN}$, $C_{NS}$ and
$C_{LN}$ are related to the imaginary parts of interference terms
and are therefore zero to lowest order of perturbative QCD\@. To
higher orders they may be non-zero. In Fig.\ \ref{fig:spin} we
show results for some of the spin correlations. Since these
predictions extend to c.m.s. scattering angles up to $90^\circ$
the $\beta$ dependence of the spin correlations has been fully
taken into account (beyond linear order) in the numerical
evaluation.

%%%%%%%%%%%%%%%%%%%%%%%%%%%%%%%%%%%%%%%%%%%%%%%%%%%%%%%%%%%%%%%%%%%%%
\section{Summary}
\label{sec:summary}
%%%%%%%%%%%%%%%%%%%%%%%%%%%%%%%%%%%%%%%%%%%%%%%%%%%%%%%%%%%%%%%%%%%%%
We have analyzed the exclusive production of $ \Lambda_c^+
\bar{\Lambda}{}_c^-$ pairs in proton-antiproton collisions. We
have argued that this process is mediated by collinear emission of
soft $u$ ($\bar{u}$) quarks from the (anti-)proton, a hard
scattering $u\bar{u}\to c\bar{c}$ and subsequently a collinear
absorption of $c$ ($\bar{c}$) quarks by the remainders of the
(anti-)proton turning them into the final $\Lambda_c^+$
($\bar{\Lambda}_c^-$). The hard subprocess is calculable within
perturbative QCD to a given order. As we showed the hadronic
matrix elements that describe the $p\to\Lambda_c^+$ ($\bar{p}\to
\bar{\Lambda}_c^-$) transition, can be parameterized in terms of
GPDs. The process amplitudes thus factorize into a product of hard
subprocess amplitudes and $1/\xb$ moments of GPDs, i.e.\
generalized form factors as, for instance, in wide-angle Compton
scattering.

To model the GPDs we employ the representation of the GPDs as
overlaps of LCWFs for the involved baryons. The LCWFs for the
valence Fock states of the baryons are taken from the literature.
Although, in general, the restriction to the valence Fock states
is insufficient for modelling the GPDs as a function of three
variables $\xb$, $\xi$ and $t$, it is expected to be a reasonable
approximation for our case. As is predicted by the HQET, the
$\Lambda_c$ valence Fock state is dominated by the simple
configuration where the $c$ quark carries the helicity of the
$\Lambda_c$ while the light quarks are coupled in a spin and
isospin zero state with only little admixture of other
configurations. It is also expected that higher Fock states play
only a minor role for the $\Lambda_c$. A possible more complicated
structure of the proton involving, for instance, parton orbital
angular momentum or higher Fock states, is irrelevant since its
overlap with the simple $\Lambda_c$ LCWF is zero. We therefore
achieve rather robust model GPDs with only a few parameters, like
the transverse size parameters or the normalizations of the LCWFs
which can be adjusted within certain ranges without loosing their
physical interpretation. With the GPDs at hand we are in the
position to evaluate the form factors and, subsequently, to
predict differential and integrated cross section as well as a
number of spin correlation parameters. We emphasize that the
prediction of the integrated cross section achieved here does not
differ much from a previous estimate in a model that bears
resemblance with the handbag approach \ci{quadder}.

We also note that the $u\to c$ transition GPDs do not only form the
soft physics input for the handbag contribution to  the process
$p\bar{p}\to \Lambda_c^+ \bar{\Lambda}{}_c^-$ but allows also to
calculate the $\Lambda_c^+\to p$ transition form factors. Another
possible application is electroproduction of charmed mesons (e.g.\
$ep\to e'D(D^*)\Lambda_c^+$). Extensions to other charmed baryons
and to calculation of $\als$ corrections to the subprocess is left
to a forthcoming paper.
%%%%%%%%%%%%%%%%%%%%%%%%%%%%%%%%%%%%%%%%%%%%%%%%%%%%%%%%%%%%%%%%%%%%%%%%%%%%%%
\section*{Acknowledgements}
We thank M. Diehl and P. Mulders for discussions. A.T.G.
acknowledges the support of the \lq\lq Fonds zur F\"orderung der
wissenschaftlichen Forschung in \"Os\-ter\-reich\rq\rq\ (FWF DK
W1203-N16).

%%%%%%%%%%%%%%%%%%%%%%%%%%%%%%%%%%%%%%%%%%%%%%%%%%%%%%%%%%%%%%%%%%%%%%%%%%%%%
\vspace*{0.05\tw} \noindent{\bf \LARGE Appendices}
%%%%%%%%%%%%%%%%%%%%%%%%%%%%%%%%%%%%%%%%%%%%%%%%%%%%%%%%%%%%%%%%%%%%%%%%%%%%
\begin{appendix}
%%%%%%%%%%%%%%%%%%%%%%%%%%%%%%%%%%%%%%%%%%%%%%%%%%%%%%%%%%%%%%%%%%%%%%%%%%%%
\section{Kinematics} \label{sec:kinematics}
%%%%%%%%%%%%%%%%%%%%%%%%%%%%%%%%%%%%%%%%%%%%%%%%%%%%%%%%%%%%%%%%%%%%%%%%%%%%
The momenta and light-cone helicities of the incoming proton and
antiproton are denoted by $p$, $\mu$ and $q$, $\nu$, those of the
outgoing $\Lambda_c^+$ and $\bar{\Lambda}{}_c^-$  by $p'$, $\mu'$,
and $q'$, $\nu'$, respectively. The mass of the proton is denoted
by $m$ that of the $\Lambda_c^+$ by $M$. We work in a center of
mass frame where the baryon's momenta in light-cone coordinates
are parameterized as
%(see Fig.\ \ref{fig:frame})
%
\ba p&=&\left[(1+\xi)\bar{p}^+,\,
\frac{m^2+\vdd/4}{2(1+\xi)\bar{p}^+},\,-\frac{\vd}2 \right ], \,
p^\prime=\left[(1-\xi )\bar{p}^+,\,
\frac{M^2+\vdd/4}{2\bar{p}^+(1-\xi)},
                             \,\phantom{-}\frac{\vd}2\right ],\nn\\[0.1em]
q&=&\left[\frac{m^2+\vdd/4}{2(1+\xi)\bar{p}^+},\,(1+\xi)\bar{p}^+,\,\phantom{-}\frac{\vd}2\right],
\, q^\prime=\left[\frac{M^2+\vdd/4}{2\bar{p}^+(1-\xi)},\,(1-\xi
)\bar{p}^+,\, -\frac{\vd}2 \right]. \label{def-momenta-ji} \ea
It is convenient to introduce sum and differences of the baryon
momenta
\be P\=2 \bar{p}\= p+p^\prime\,, \qquad Q\=q+q^\prime\,, \qquad
\Delta=p^\prime-p=q-q^\prime\,. \label{sum-and-diff} \ee
The momentum transfer $\Delta$ is expressed by
\be \Delta \=\Big[-2\xi\bar{p}^+\,,
  \frac{M^2(1+\xi)-m^2(1-\xi)+\xi\vdd/2}{2\bar{p}^+(1-\xi^2)}\,,
  \vd\Big]\,.
\ee
The three-components of the c.m.s.\ momenta of the incoming (anti)
proton, $(-)p_3$, and the outgoing (anti) $\Lambda_c$,
$(-)p^\prime_3$, read
\be p_3\= \frac12\sqrt{s}\sqrt{\Lambda_m^2-\vdd/s}\,, \quad \mid
p^\prime_3\mid\= \frac12\sqrt{s}\sqrt{\Lambda_M^2-\vdd/s}\,, \ee
where $s=(p+q)^2=(p^\prime+q^\prime)^2$ is the usual Mandelstam
energy variable. We also introduce the abbreviation
\be \Lambda_i \= \sqrt{1-4m_i^2/s}  \label{eq:lambdami}\ee
in a more generic notation ($m_i$ stands for $m$, $M$, or the
charm-quark mass $m_c$; for $m_i=m_c$ Mandelstam $s$ has to be
replaced by $\hat{s}$). While the three-component of the incoming
proton's momentum is always positive that of the outgoing
$\Lambda_c$ can become negative for large c.m.s.\ scattering
angles because of the unequal-mass kinematics. This change of
signs occurs at
\be \vdd{}_{\rm max} \= s \Lambda^2_M\,. \ee
Thus, due to the unequal-mass kinematics, $\vdd$ is zero for
forward scattering, reaches the maximal value and decreases again
towards zero for backward scattering. Skewness, defined as a ratio
of light-cone plus-components of baryon momenta,
$\xi=(p^+-p^{\prime +})/(p^++p^{\prime +})$, can be expressed as
\ba \xi&=&\frac{\sqrt{\Lambda_m^2-\vdd/s} -{\rm
sign}(p^\prime_3)\sqrt{\Lambda_M^2-\vdd/s}}
   {2+ \sqrt{\Lambda_m^2-\vdd/s} +{\rm sign}(p^\prime_3)
     \sqrt{\Lambda_M^2-\vdd/s}}\nn\\
  & =& \frac{\Lambda_m^2 - \Lambda_M^2}{\sqrt{\Lambda_m^2 +
       \Lambda_M^2 + 2 \Lambda_m\Lambda_M\cos\theta}} \,
      \frac{1}{2 + \sqrt{\Lambda_{m}^2 + \Lambda_{M}^2 + 2 \Lambda_{m}\Lambda_{M}\cos\theta}}\nn\\
  & \to& \frac{M^2-m^2}{2s}\,\Big[1+\frac32\frac{M^2+m^2+\vdd/2}{s}\Big]
       \quad{\rm for\;} s\to\infty\, ,
\label{skew} \ea
where $\theta$ is the c.m.s.\ scattering angle. In a similar
fashion can the other parameter occurring in the definition of the
baryon momenta \req{def-momenta-ji} be written as
\ba \bar{p}^+ &=& \frac14\sqrt{\frac{s}{2}} \Big[2+
\sqrt{\Lambda_m^2-\vdd/s} +{\rm sign}(p^\prime_3)
\sqrt{\Lambda_M^2-\vdd/s}\Big]\nn\\
       &=& \frac14 \sqrt{\frac{s}{2}} \Big[2 + \sqrt{\Lambda_m^2 + \Lambda_M^2
       + 2\Lambda_m\Lambda_M \cos\theta } \Big]\,.
\ea
As a consequence of the unequal mass kinematics $\xi$ cannot
become zero. For $p'_3\geq 0$, however, the skewness is fairly
small and tends to zero for $s\to\infty$.

The squared invariant momentum transfer is given by
\ba t&=&\Delta^2\= -\frac{\vdd}{1-\xi^2} -
\frac{2\xi}{1-\xi^2}\Big[(1+\xi) M^2
-(1-\xi) m^2\Big] \nn\\
 &=&-\frac{\vdd}{2}
         -\frac{s}{4}\Big[\Lambda_m^2+\Lambda_M^2
    -2{\rm
      sign}(p^\prime_3)\sqrt{\Lambda_m^2-\vdd/s}\sqrt{\Lambda_M^2-\vdd/s}\Big]\nn\\
 &=& -\frac{s}{4} \Big[\Lambda_m^2 + \Lambda_M^2 - 2 \Lambda_m\Lambda_M\cos\theta \Big]\,.
\label{def:t}
\ea
It cannot become zero for forward scattering but acquires the value
($\xi_0=\xi(\vdd=0,p'_3\geq0)$):
\ba
t_{0}\=t(\vdd=0, p^\prime_3\geq 0) &=& -\frac{2\xi_0}{1-\xi_0^2}\Big[(1+\xi_0)M^2-(1-\xi_0)m^2\Big]\nn\\
              &=& -\frac{s}4\,(\Lambda_m-\Lambda_M)^2\,.
\label{min-t}
\ea
It is convenient to introduce a variable $t'$ that vanishes for forward
scattering. It is defined by
\ba
t^\prime&=&t-t_0\=-\frac{\vdd}{2}-\frac{s}{2}\Big[\Lambda_m\Lambda_M
 - {\rm sign}(p'_3)\sqrt{\Lambda_m^2-\vdd/s}\sqrt{\Lambda_M^2-\vdd/s}\Big]\nn\\
   &=& -\frac{s}{2} \Lambda_{m} \Lambda_{M} \Big[1 - \cos\theta \Big]\,.
\label{def:tprime}
\ea
Solving \req{def:tprime} for $\vdd$, one finds
\be
\vdd\=-t^\prime\frac{s\Lambda_m\Lambda_M+t^\prime}{s/4(\Lambda_m+\Lambda_M)^2+t^\prime}
   \= s \frac{\Lambda_m^2 \Lambda_M^2 \sin^2\theta}
            {\Lambda_m^2 + \Lambda_M^2 + 2 \Lambda_m\Lambda_M\cos\theta}\,.
\ee
Sine and cosine of half the c.m.s.\ scattering angle $\theta$ are
connected with  the Mandelstam variables by
\ba \sin^2{\theta/2} &=& \frac{1-\cos{\theta}}{2} \=
  \frac{t_0-t}{s\Lambda_m\Lambda_M}\,,\nn\\
\cos^2{\theta/2} &=& \frac{1+\cos{\theta}}{2} \=
  \frac{u_1-u}{s\Lambda_m\Lambda_M}\,,
\ea
where
\be u_1\=u(\vdd=0,p^\prime_3\leq 0)\=
-\frac{s}4\,(\Lambda_m-\Lambda_M)^2\,. \ee
For forward scattering $u$ reads
\be u_0\=u(\vdd=0,p^\prime_3\geq 0)\=
-\frac{s}4\,(\Lambda_m+\Lambda_M)^2\,. \ee
%

%%%%%%%%%%%%%%%%%%%%%%%%%%%%%%%%%%%%%%%%%%%%%%%%%%%%%%%%%%%%%%%%%%%%%%%%%%%%%
\section{The $u\to c$ generalized parton distributions}
\label{app-B}
%%%%%%%%%%%%%%%%%%%%%%%%%%%%%%%%%%%%%%%%%%%%%%%%%%%%%%%%%%%%%%%%%%%%%%%%%%%%%
The vector current of bilocal quark field operators is defined as
\be V^\mu(-z^-/2,z^-/2) \= \bar{\Psi}^c(-z^-/2)\gamma^\mu
\Psi^u(z^-/2)
              - \bar{\Psi}^u(z^-/2)\gamma^\mu  \Psi^c(-z^-/2)\, .
\label{def:V}
\ee
The Fourier transform of the plus component of its transition
matrix element
\be {\cal H}_{\mu^\prime\mu}^{\,cu} \= \bar{p}^+ \int
\frac{dz^-}{2\pi} e^{\imath \xb\bar{p}^+z^-}
                \langle \Lambda_c^+;p^\prime \mu^\prime\mid V^+(-z^-/2,z^-/2)\mid p; p\mu\rangle\,,
\label{V-ME}
\ee
can be decomposed into five covariant structures
\ba &&\frac{P^+}{M+m}\bar{u}(p^\prime)u(p)\,, \quad
\frac{\Delta^+}{M+m}\bar{u}(p^\prime)u(p) \,, \quad
\bar{u}(p^\prime)\gamma^+ u(p)\,,  \nn\\
&&\bar{u}(p^\prime)\frac{\imath\sigma^{+\nu}\Delta_\nu}{M+m} u(p)\,,
     \quad \bar{u}(p^\prime)\frac{\imath\sigma^{+\nu}P_\nu}{M+m} u(p)\,.
\ea
The combinations of momenta are defined in \req{sum-and-diff}.
Helicity labels are omitted wherever it can be done without loss
legibility. Only two of the covariant structures are independent
since we have the familiar Gordon decomposition and analogous
relations at disposal
\ba P^\mu \bar{u}(p^\prime)u(p) &=&
(M+m)\bar{u}(p^\prime)\gamma^\mu u(p) -\imath
\bar{u}(p^\prime) \sigma^{\mu\nu}\Delta_\nu u(p)\,,\nn\\
\Delta^\mu \bar{u}(p^\prime)u(p) &=& (M-m)
\bar{u}(p^\prime)\gamma^\mu u(p) -\imath \bar{u}(p^\prime)
\sigma^{\mu\nu}P_\nu u(p)\,. \label{gordon-vector} \ea
In addition we have for the light-cone projections the relation
\be \Delta^+\= -\xi P^+\,, \label{projection} \ee
see \req{def-momenta-ji} and \req{sum-and-diff}. The two
independent GPDs are chosen in analogy to the flavor-diagonal case
(cf.\ \ci{bel05})
\be {\cal H}^{\,cu}_{\mu^\prime\mu} \=
\bar{u}(p^\prime,\mu^\prime)\left[H^{cu}(\xb,\xi,t)\,\gamma^+
                     + E^{cu}(\xb,\xi,t)\,\frac{\imath\sigma^{+\nu}\Delta_\nu}{M+m}\right]
                     u(p,\mu)\,.
\label{V-FF}
\ee

In the unequal mass-case the matrix elements of the local vector
current describing the electroweak $p\to \Lambda_c^+$ transition
can be decomposed into three covariant structures for which one
may choose
\be \langle \Lambda_c^+;p^\prime \mu^\prime\mid V^\mu(0,0) \mid
p;p \mu\rangle \= \bar{u}(p^\prime,\mu^\prime) \Big[
   F_V \gamma^\mu\; +\; F_T \frac{\imath \sigma^{\mu\nu}\Delta_\nu}{m+M}\;
                  +\; F_3 \frac{\Delta^\mu}{M+m}\Big] u(p,\mu)\,.
\label{vector-current}
\ee
The four-vectors $\Delta^\mu$ and $P^\mu$ are independent; only their
plus components are proportional to each other.

Let us integrate \req{V-ME} and \req{V-FF} with respect to
$\bar{x}$, which reduces the bilocal matrix element to a local one
that describes the weak transition form factors for which we
insert the form factor decomposition \req{vector-current}
\ba \int_{-1}^1 d\xb {\cal H}^{\,cu}_{\mu^\prime\mu} &=&
           \langle \Lambda_c^+;p^\prime\mu^\prime\mid V^+(0,0) \mid p;p\mu\rangle \nn\\
    && \hspace*{-0.15\tw} =\int_{-1}^1 d\xb H^{cu}(\xb,\xi,t)\, \bar{u}(p^\prime,\mu^\prime)
    \gamma^+u(p,\mu)
             + \int_{-1}^1 d\xb E^{cu}(\xb,\xi,t)\,
               \bar{u}(p^\prime,\mu^\prime)\frac{\imath\sigma^{+\nu}\Delta_\nu}{M+m}
               u(p,\mu) \nn\\
         & & \hspace*{-0.15\tw} = \bar{u}(p^\prime,\mu^\prime) \left[ F_V(t)\, \gamma^+
         \; +\; F_T(t)\, \frac{\imath \sigma^{+\nu}\Delta_\nu}{m+M}\;
              +\; F_3(t)\, \frac{\Delta^+}{M+m} \right]u(p,\mu)\,.
\ea
Using again the Gordon decomposition \req{gordon-vector} for the term
$\propto F_3$, we find the following sum rules
\ba
\int_{-1}^1 d\xb H^{cu}(\xb,\xi,t) &=& F_V(t) - \xi F_3(t)\,, \nn\\
\int_{-1}^1 d\xb E^{cu}(\xb,\xi,t) &=& F_T(t) + \xi F_3(t)\,.
\label{vector-sum-rules}
\ea
In the limit of equal hadron masses conservation of the vector current
requires $F_3=0$. To lowest order of the electroweak theory
the weak transition form factors are zero; this holds in particular at the
scale of $M_Z$. For lower scales, like $M$, there are effective operators
that mediate $u\to c$ transitions through finite Wilson coefficients.

The lower vertex is treated analogously. It is related to the upper
one by charge conjugation. It is then easy to show that
\be \langle \Lambda_c^+\mid V^+(-z/2,z/2)\mid p\rangle
           \= - \langle \bar{\Lambda}_c^-\mid V^+(-z/2,z/2)\mid \bar{p}\rangle\,.
\ee
Taking the Fourier transform of this matrix element in analogy to
\req{V-ME}, making the replacements
\be p\to q\,, \quad p^\prime\to q^\prime\,, \quad \Delta\to
-\Delta\,, \quad \gamma^+\to\gamma^-\,, \ee
and rewriting the right-hand side of \req{V-FF} with the help of
the behavior of the Dirac spinors under charge conjugation one
finds for the lower vertex
\ba {\cal H}^{\,\overline{cu}}_{\nu^\prime \nu} &=&
  \bar{q}^-\int \frac{dz^+}{2\pi}e^{\imath \xb\bar{q}^-z^+}
    \langle \bar{\Lambda}_c^-;q^\prime\nu^\prime\mid V^-(-z^+/2,z^+/2)\mid\bar{p};q\nu\rangle \nn\\
        &=& \bar{v}(q,\nu)\left[ H^{cu}(\xb,\xi,t)\,\gamma^- -
          E^{cu}(\xb,\xi,t)\,\frac{\imath\sigma^{-\nu}\Delta_\nu}{M+m}\right]
      v(q^\prime,\nu^\prime)\,.
\ea
The lowest moment of ${\cal H}_{\nu'\nu}^{\overline{cu}}$ leads to
the vector-current matrix element
\ba \int_{-1}^1 d\xb {\cal H}^{\,\overline{cu}}_{\nu^\prime \nu}
&=&
           \langle \bar{\Lambda}_c^-;q^\prime\nu^\prime\mid V^-(0,0)
           \mid\bar{p};q\nu\rangle \nn\\
      && \hspace*{-0.15\tw}  =\int_{-1}^1 d\xb H^{cu}(\xb,\xi,t)\,
      \bar{v}(q,\nu)\gamma^-v(q^\prime,\nu^\prime)
             - \int_{-1}^1 d\xb E^{cu}(\xb,\xi,t)\,
               \bar{v}(q,\nu)\frac{\imath\sigma^{-\nu}\Delta_\nu}{M+m}
               v(q^\prime,\nu^\prime) \nn\\
      && \hspace*{-0.15\tw}  =\bar{v}(q,\nu)\,\left[ F_V(t)\,\gamma^-
              -\; F_T(t)\,\frac{\imath \sigma^{-\nu}\Delta_\nu}{m+M}\;
              -\; F_3(t)\, \frac{\Delta^-}{M+m}\right]\,v(q^\prime,\nu^\prime)\,.
\ea
Using
\be \Delta^-\= \xi Q^-\,, \ee
and the Gordon decomposition
\be Q^\mu \bar{v}(q)v(q^\prime) \= - (M+m) \bar{v}(q)\gamma^\mu
v(q^\prime) -\imath \bar{v}(q) \sigma^{\mu\nu}\Delta_\nu
v(q^\prime)\,, \ee
we find exactly the same sum rules as in \req{vector-sum-rules}.

Next we proceed analogously for the axial vector transition matrix
element:
\be A^\mu(-z^-/2,z^-/2) \= \bar{\Psi}^c(-z^-/2)\gamma^\mu\gamma_5
\Psi^u(z^-/2)
              - \bar{\Psi}^u(z^-/2)\gamma^\mu\gamma_5
              \Psi^c(-z^-/2)\, ,
\ee
\be
\widetilde{\cal H}_{\mu'\mu}^{\,cu} \= \bar{p}^+ \int \frac{dz^-}{2\pi}\,e^{\imath \xb\bar{p}^+z^-}
              \langle \Lambda_c^+; p^\prime\mu^\prime\mid A^+(-z^-/2,z^-/2)\mid p;p\mu\rangle\,,
\label{A-ME}
\ee
can again be decomposed into five covariant structures
\ba &&\frac{P^+}{M+m}\bar{u}(p^\prime)\gamma_5u\,, \quad
\frac{\Delta^+}{M+m}\bar{u}(p^\prime)\gamma_5u\,, \quad
\bar{u}(p^\prime)\gamma^+\gamma_5 u\,, \nn\\
&&\bar{u}(p^\prime)\frac{\imath\sigma^{+\nu}\Delta_\nu\gamma_5}{M+m} u\,,
     \quad \bar{u}(p^\prime)\frac{\imath\sigma^{+\nu}P_\nu\gamma_5}{M+m} u\,.
\ea
We also have the two relations
\ba
\bar{u}(p^\prime)\imath\sigma^{\mu\nu}P_\nu\gamma_5 u(p) &=&
                  (M+m) \bar{u}(p^\prime) \gamma^\mu\gamma_5 u(p) - \Delta^\mu
           \bar{u}(p^\prime)\gamma_5 u(p)\,,\nn\\
\bar{u}(p^\prime) \imath\sigma^{\mu\nu}\Delta_\nu\gamma_5 u(p) &=&
                  (M-m) \bar{u}(p^\prime) \gamma^\mu\gamma_5 u(p) - P^+
           \bar{u}(p^\prime)\gamma_5 u(p)\,,
\label{gordon-axial} \ea
and the projection \req{projection}. The two independent GPDs are
chosen in analogy to the flavor-diagonal case
\be \widetilde{{\cal H}}^{\,cu}_{\mu^\prime \mu} \=
\bar{u}(p^\prime,\mu^\prime) \left[
          \widetilde{H}^{cu}(\xb,\xi,t)\,\gamma^+
            + \widetilde{E}^{cu}(\xb,\xi,t)\,
                   \frac{\Delta^+}{M+m} \right]\gamma_5 u(p,\mu)\,.
\label{A-FF} \ee
We define the general local axial-vector current matrix element as
\be \langle \Lambda_c^+;p^\prime\mu^\prime\mid A^\mu(0,0) \mid
p;p\mu\rangle \= \bar{u}(p^\prime,\mu^\prime) \Big[
   G_A \gamma^\mu\; +\; G_P \frac{\Delta^\mu}{m+M}\;
                  +\; G_3 \frac{P^\mu}{M+m}\Big]\gamma_5 u(p,\mu)\,.
\label{axial-current} \ee
This may be compared with a decomposition which is useful for
applications of the HQET \cite{koerner}. Integration of the GPD
with respect to $\xb$ leads to
\ba \int_{-1}^1 d\xb \widetilde{\cal H}_{\mu^\prime\mu}^{\,cu} &=&
           \langle \Lambda_c^+;p^\prime\mu^\prime \mid A^\mu(0,0) \mid p; p\mu\rangle \nn\\
    && \hspace*{-0.15\tw}    = \int_{-1}^1 d\xb \widetilde{H}^{cu}(\xb,\xi,t)\,
       \bar{u}(p^\prime,\mu^\prime)\gamma^+\gamma_5 u(p,\mu)
             + \int_{-1}^1 d\xb \widetilde{E}^{cu}(\xb,\xi,t)\,
               \frac{\Delta^+}{M+m}\bar{u}(p^\prime,\mu^\prime)\gamma_5 u(p,\mu) \nn\\
     &&  \hspace*{-0.15\tw}   = \bar{u}(p^\prime,\mu^\prime) \left[ G_A(t)\, \gamma^+ \;
            + G_P(t) \frac{\Delta^+}{M+m}
            + G_3(t) \frac{P^+}{M+m}  \right]\gamma_5  u(p,\mu)\, .
\ea
Using \req{projection} we finally obtain the sum rules
\ba
\int_{-1}^1 d\xb \widetilde{H}^{cu}(\xb,\xi,t) &=& G_A(t) \,, \nn\\
\int_{-1}^1 d\xb \widetilde{E}^{cu}(\xb,\xi,t) &=& G_P(t) -
          \frac1{\xi}G_3(t)\,.
\label{axial-sum-rules}
\ea

Applying charge conjugation
\be \langle \Lambda_c^+\mid A^+(-z/2,z/2)\mid p\rangle \=
          - \langle \bar{\Lambda}_c^-\mid A^+(-z/2,z/2)\mid\bar{p}\rangle\,,
\ee
one obtains the parameterization of the lower vertex
\ba \widetilde{{\cal H}}^{\,\overline{cu}}_{\nu^\prime \nu} &=&
     \bar{q}^-\int \frac{dz^+}{2\pi} e^{\imath \xb \bar{q}^-z^+}
     \langle \bar{\Lambda}^-_c;q^\prime \nu^\prime\mid A^-(-z^+/2,z^+/2
     \mid\bar{p}; q \nu \rangle \nn\\
     &=& \bar{v}(q, \nu)\left[\widetilde{H}^{cu}\,  \gamma^-\gamma_5
         - \widetilde{E}^{cu}\, \frac{\Delta^-\gamma_5}{M+m} \right]
     v(q^\prime,\nu^\prime)\,.
\ea
The lowest moment becomes
\ba \int_{-1}^1 d\xb\widetilde{{\cal H}}^{\,\overline{cu}}&=&
      \langle \bar{\Lambda}_c^-; q^\prime\nu^\prime\mid A^+(0,0)\mid \bar{p} ;q\nu\rangle \nn\\
   && \hspace*{-0.15\tw} = +\int_{-1}^1 d\xb \widetilde{H}^{cu}
   \bar{v}(q,\nu)\gamma^-\gamma_5 v(q^\prime,\nu^\prime)
                - \int_{-1}^1 d\xb \widetilde{E}^{cu}
        \bar{v}(q,\nu)\frac{\Delta^-\gamma_5}{M+m}v(q^\prime,\nu^\prime)\nn\\
   &&  \hspace*{-0.15\tw}  = \bar{v}(q,\nu) \left[G_A(t) \gamma^-
        -G_P(t)\frac{\Delta^-}{M+m}
                +G_3(t)\frac{Q^-}{M+m}\right]\gamma_5   v(q^\prime,\nu^\prime)\,.
\ea
One reads again off the sum rules \req{axial-sum-rules}.

In an analogous way the helicity-flip GPDs can be written as (for
definitions see Ref.\ \cite{diehl01}):
\ba {\cal H}^{Tcu}_{j\mu^\prime\mu}&=&
\bar{p}^+ \int \frac{dz^-}{2\pi}\,e^{\imath \xb_1\bar{p}^+z^-} \nn\\
          &\times&               \langle\Lambda_c^+; p^\prime\mu^\prime\mid
           \bar{\Psi}^c(-z^-/2)\imath \sigma^{+j}  \Psi^u(z^-/2) -
         \bar{\Psi}^u(z^-/2)\imath\sigma^{+j}  \Psi^c(-z^-/2) \mid p; p \mu\rangle\,  \nn\\
         &=& \bar{u}(p^\prime,\mu^\prime)\Big[H^{cu}_T\imath\sigma^{+j}
                     + \widetilde{H}^{cu}_T \frac{P^+\Delta^j-\Delta^+P^j}{2mM} \nn\\
         &+& E^{cu}_T \frac{\gamma^+\Delta^j-\Delta^+\gamma^j}{M+m}
                     + \widetilde{E}^{cu}_T\frac{\gamma^+P^j-P^+\gamma^j}{M+m} \Big]
                                       u(p,\mu)\,.
\label{eq:GPD-flip}
\ea
Charge conjugation symmetry leads to
\ba {\cal H}^{T\overline{cu}}_{j\, \nu^\prime\nu}&=&
\bar{q}^- \int \frac{dz^+}{2\pi}\,e^{\imath \xb\bar{q}^-z^+} \nn\\
           &\times&              \langle \bar{\Lambda}_c^-; q^\prime \nu^\prime\mid
           \bar{\Psi}^c(-z^+/2)\imath \sigma^{-j}  \Psi^u(z^+/2) -
         \bar{\Psi}^u(z^+/2)\imath\sigma^{-j}  \Psi^c(-z^+/2) \mid \bar{p};q \nu\rangle\,  \nn\\
         &=& \bar{v}(q,\nu)\Big[-H^{cu}_T\imath\sigma^{-j}
                     - \widetilde{H}^{cu}_T \frac{Q^-\Delta^j-\Delta^-Q^j}{2mM} \nn\\
         &+& E^{cu}_T \frac{\gamma^-\Delta^j-\Delta^-\gamma^j}{M+m}
                     - \widetilde{E}^{cu}_T\frac{\gamma^-Q^j-Q^-\gamma^j}{M+m} \Big]
                                       v(q^\prime,\nu^\prime)\,.
\label{eq:GPD-flip-anti}
\ea
With the relations
\be \sigma^{\pm 1}\gamma_5\=\mp\imath \sigma^{\pm 2}\,, \qquad
\sigma^{\pm 2}\gamma_5\=\pm \imath\sigma^{\pm 1}\,, \ee
%
%where the convention is $\varepsilon^{0123}=1$, the form factor decomposition
a combination of $\sigma^{+1}$  and $\sigma^{+2}$ matrix elements
\req{eq:GPD-flip} and \req{eq:GPD-flip-anti} can be reexpressed as
\be
\frac12\big[\imath\sigma^{\pm1}\pm2\lambda_1\sigma^{\pm2}\big]\=
                \frac12\imath\sigma^{\pm1}\big[1+2\lambda_1\gamma_5\big]\,.
\label{ab-a5}
\ee

\end{appendix}
%%%%%%%%%%%%%%%%%%%%%%%%%%%%%%%%%%%%%%%%%%%%%%%%%%%%%%%%%%%%%%%%%%%%%%%%%%%

\end{document}